\documentclass[a4paper,fleqn,usenatbib]{mnras}
\usepackage{newtxtext,newtxmath}

\usepackage[T1]{fontenc}
\usepackage{ae,aecompl}

\usepackage[english]{babel}
\usepackage{amsmath,amsfonts,amssymb,textcomp}
\usepackage{amssymb}
\usepackage{pdflscape}
\usepackage{multirow}
\usepackage{varioref}
\usepackage{hyperref}
\usepackage{psfig}
\usepackage{longtable}
\usepackage{graphicx,graphics,subfigure}
\usepackage[usenames,dvipsnames]{xcolor}
\hypersetup{backref,
colorlinks=true,citecolor=blue, urlcolor=magenta}

\labelformat{equation}{\textup{(#1)}}

\labelformat{Fig.}{\textup{(#1)}}

\graphicspath{{imgs/}}

\title[LoCuSS: environment, star formation and dust in A1758]{LoCuSS: exploring the connection between local environment, star formation and dust mass in Abell 1758}
\author[M. Bianconi et al.]{
M. Bianconi$^{1,}\thanks{mbianconi@star.sr.bham.ac.uk}$, G. P. Smith$^{1}$, C. P. Haines$^{2,1}$, S. L. McGee$^{1}$, A. Finoguenov$^{3}$,
\newauthor \hspace{0.12cm}E. Egami$^{4}$\\\\
$^{1}$School of Physics and Astronomy, University of Birmingham, Edgbaston, Birmingham, B15 2TT, UK\\
$^{2}$Instituto de Astronom\'ia y Ciencias Planetarias de Atacama, Universidad de Atacama, Copayapu 485, Copiap\'o, Chile\\
$^{3}$Department of Physics, University of Helsinki, Gustaf  H\"{a}llstr\"{o}min katu 2a, FI-0014 Helsinki, Finland\\
$^{4}$Steward Observatory, University of Arizona, 933 North Cherry Avenue, Tucson, AZ 85721, USA
}

\date{Accepted XXX. Received YYY; in original form ZZZ}

\pubyear{2019}

\begin{document}
\label{firstpage}
\pagerange{\pageref{firstpage}--\pageref{lastpage}}
\maketitle

\begin{abstract}
We explore the connection between dust and star formation, in the context of environmental effects on galaxy evolution. In particular, we exploit the susceptibility of dust to external processes to assess the influence of dense environment on star-forming galaxies. We have selected cluster Abell 1758 from the Local Cluster Substructure Survey (LoCuSS). Its complex dynamical state is an ideal test-bench to track dust removal and destruction in galaxies due to merger and accretion shocks. We present a systematic panchromatic study (from $\rm 0.15\mu$m with GALEX to $\rm 500\mu$m with \textit{Herschel}) of spectroscopically confirmed star-forming cluster galaxies at intermediate redshift. We observe that the main subclusters (A1758N and A1758S) belong to two separate large-scale structures, with no overlapping galaxy members. Star-forming cluster members are found preferentially outside cluster central regions, and are not isotropically distributed. Rather, these galaxies appear being funneled towards the main subclusters along separate accretion  paths. Additionally, we present the first study of dust-to-stellar (DTS) mass ratio used as indicator for local environmental influence on galaxy evolution. Star-forming cluster members show lower mean values (32$\rm\%$ at $\rm2.4\sigma$) of DTS mass ratio and lower levels of infrared emission from birth clouds with respect to coeval star-forming field galaxies. This picture is consistent with the majority of star-forming cluster members infalling in isolation. Upon accretion, star-formation is observed to decrease and warm dust is destroyed due to heating from the intracluster medium radiation, ram-pressure stripping and merger shocks.
\end{abstract}

\begin{keywords}
galaxies: clusters: individual: Abell 1758 -- galaxies: evolution -- galaxies: star formation
\end{keywords}
\section{Introduction}
\begin{figure*}
 \centering
\includegraphics[width=0.9\linewidth, keepaspectratio]{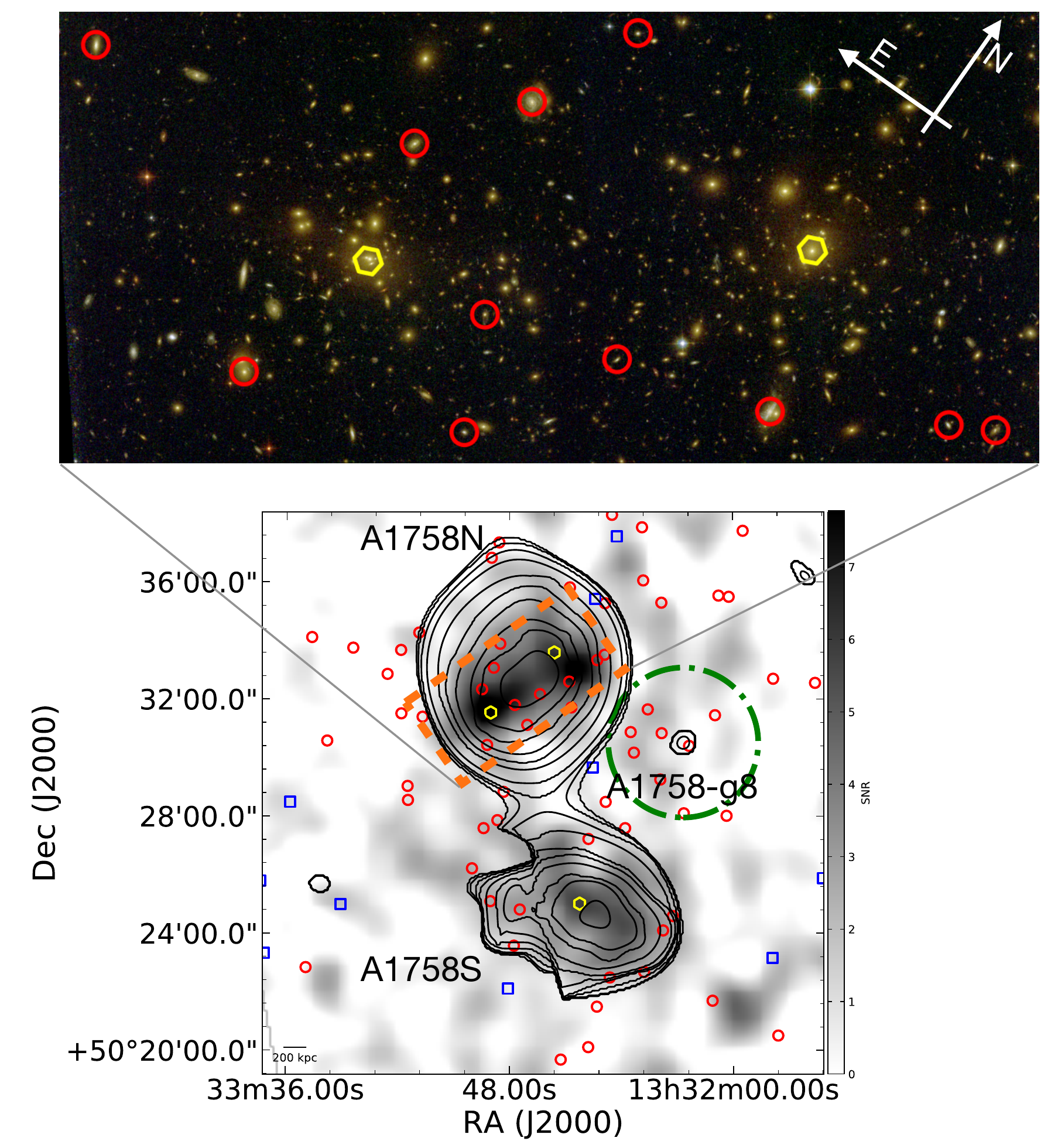}
\caption{The merging cluster A1758. Top panel: zoom-in mosaic of A1758N observed with HST-ACS in F435W, F606W and F814W filters from the RELICS survey \citep{coe19}. Bottom panel: the underlying map in shades of black shows the surface mass density  signal-to-noise (SNR) ratio based on the weak-lensing analysis of (\citealt{okabe16}, see end of Section~\ref{results}). Black contours trace the extended X-ray emission measured with \textit{XMM-Newton}, detected above a 4$\rm \sigma$ threshold in the wavelet analysis \citep{haines17}, and are logarithmically spaced between $\rm 6.3\times10^{-7}$ and $\rm 3.9\times10^{-5}\, count\,s^{-1}$. Star-forming cluster and field galaxies are plotted as red circles and blue squares, respectively. Yellow hexagons mark the brightest galaxy position of the three cluster subclumps, referred to as NW, NE and S according to their coordinates. The X-ray contours encompassed by the green dot-dashed circle (with radius equal to $\rm r_{200}$) correspond to the X-ray group A1758-g8 discovered  in \citet{haines17}.}\label{wl_maps}
 \end{figure*}

Dust plays an important role in shaping the evolution of galaxies. It acts as a catalyst for the formation of molecular gas, which accumulates in the dense and cold clouds that become the birthplace of stars (\citealt{galliano17}, for a review). Dust is also responsible for reprocessing UV radiation from newly-born stars, resulting in an extinction of light from galaxies at short wavelengths, and a re-emitting of that energy at infrared wavelengths. It is thought that dust formation occurs predominantly via the growth of grains in external layers of AGB star atmospheres and supernovae ejecta, which are later distributed into the interstellar medium by stellar winds. 
 \begin{figure}
 \centering
 \includegraphics[width=\linewidth, keepaspectratio]{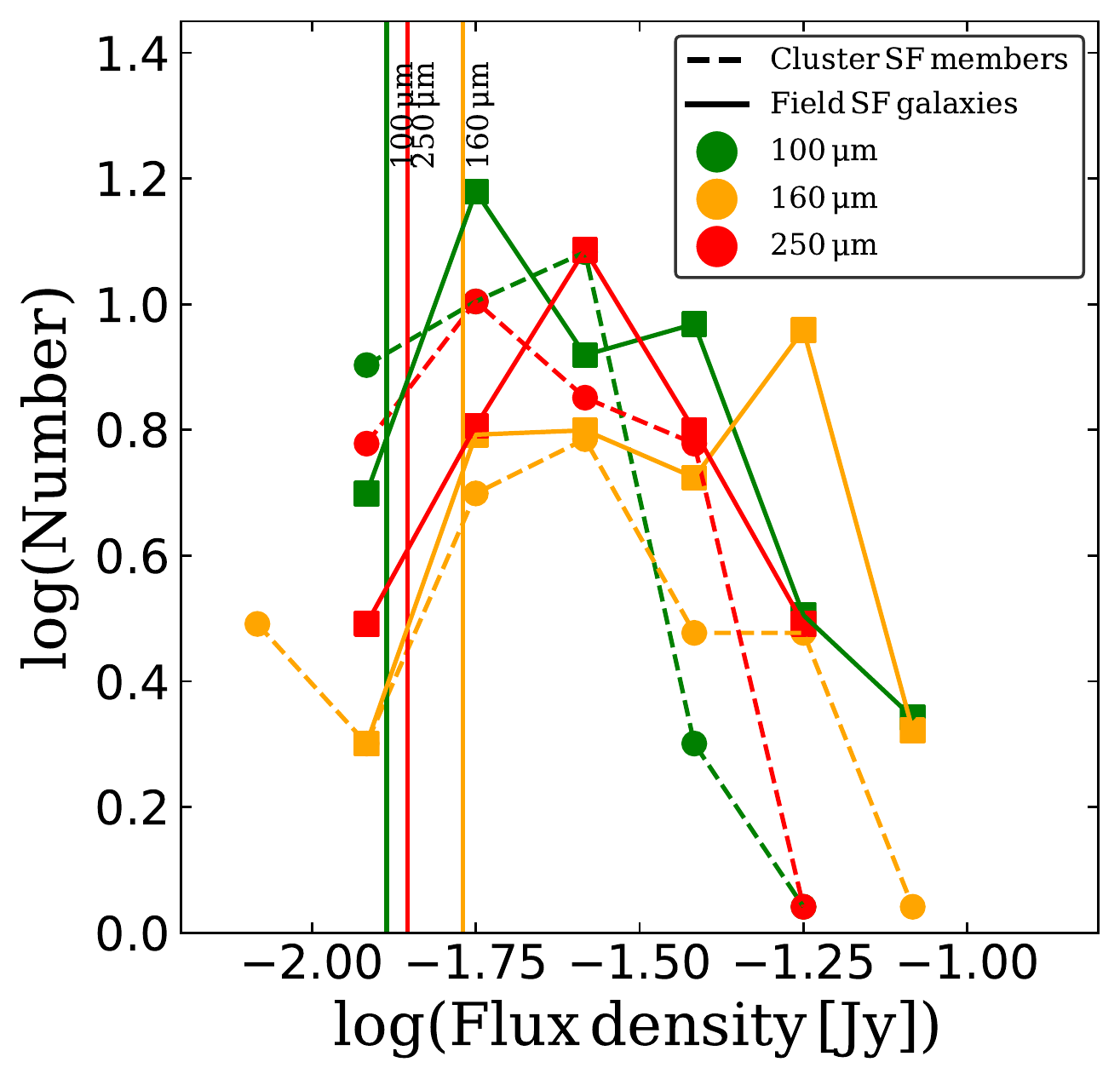}
\caption{Number of star forming galaxies detected in our far-IR observations as a function of flux density at $100$, $160$, and $250\mu{\rm m}$.  At bright fluxes, the number counts of star-forming cluster galaxies falls more steeply than the number of star-forming field galaxies, suggesting a dearth of IR-bright galaxies in clusters relative to the field. 
The vertical lines correspond to the flux limits in each infrared band. }\label{num_count}
 \end{figure}
 

\begin{table*}
\centering
\begin{tabular}{c c c c c c  c c}
\hline
Name & Centre (RA, Dec) & N &Redshift & $\rm L_{X} $ & Mass $\rm M_{200}$  & Radius $\rm r_{200}$& Velocity dispersion \\
& (RA, Dec) &  &$\rm \langle z \rangle $& $\rm (0.1-2.4keV) \, [10^{44}\, erg\,s^{-1}]$ &$\rm [10^{14}\, M_{\odot}] $&  [Mpc]&  [$\rm km\,s^{-1}$]\\
\hline
\hline
A1758N & 203.18848, 50.54294 & 176 &0.27879 $\rm \pm$ 0.0064& $\rm 7.514^{a}$ &18.21 $\rm \pm$ 3.59$\rm^{c}$& 2.77 $\rm \pm$ 0.18 & 1440 $\rm \pm$ 104\\
A1758S & 203.13729,  50.41702 & 74 &0.27386 $\rm \pm$  0.00426& $\rm 4.056^{b}$  &7.72 $\rm \pm$ 1.50 $\rm^{d}$& 1.74 $\rm \pm$ 0.11  & 1020 $\rm \pm$ 78\\
A1758-g8 & 203.04446,	50.50874 & 17 &0.27891 $\rm \pm$  0.00097& $\rm 0.041^{b}$ &0.41 $\rm \pm$ 0.07 $\rm^{d}$& 0.65 $\rm \pm$ 0.04 & 300 $\rm \pm$ 81\\ 
\hline
\end{tabular}
\caption{Summary of the principal properties of the two main subclusters, A1758N and S, and the X-ray group A1758-g8 from \citet{haines17}. From left to right, halo name, central coordinate, number of spectroscopically confirmed members, X-ray luminosity $\rm L_{X}$($\rm^{a}$ from \textit{ROSAT}, $\rm^{b}$ from \textit{XMM-Newton}, \citealt{haines17}), mass $\rm M_{200}$ ($\rm^{c}$ from the combined \textit{Chandra-XMM} analysis of \citealt{martino14}, whereas  $\rm^{d}$ is computed using the scaling relation between X-ray luminosity $\rm L_{x}$ and $\rm M_{200}$ from \citealt{leauthaud10}), radius $\rm r_{200}$ and velocity dispersion, which is estimated from the velocity distribution of member galaxies. $\rm M_{200}$ is defined as the mass contained within $\rm r_{200}$, which encompasses an overdense region presenting an average density 200 times higher than the Universe critical density at the cluster redshift $\rm \rho_{crit}(z)$, i.e. $\rm M_{200} = \frac{4}{3} \pi r_{200}^3 200 \rho_{crit}(z)$  \citep{voit2005}. }\label{table_props}
\end{table*}
As it traces the creation of galaxy's stellar content, and is mixed through the interstellar medium, measurements of the dust content are crucial for understanding why the star formation rate density of universe has declined since $z\simeq2$ and what drives the quenching of star formation. Observations have shown that while star-forming galaxies have high dust content (particularly as a fraction of their stellar mass), passive galaxies do not \citep{smith12}. This has been extended by showing that the dust mass of a galaxy directly correlates with the star formation rate (SFR), at least for galaxies in the field \citep{dacunha10}. It is not clear what happens to the dust created during star formation, such that it is no longer detected in massive and passive galaxies. It has been proposed that it is destroyed via mechanisms internal to the galaxy, such as supernovae shocks (\citealt{jones04}, for a review), or is driven out of the galaxy by an outflow or consumed less efficiently due to heating from active galactic nuclei \citep{gobat18}.  
    \begin{figure*}
 \centering

\includegraphics[width=0.38\linewidth, keepaspectratio]{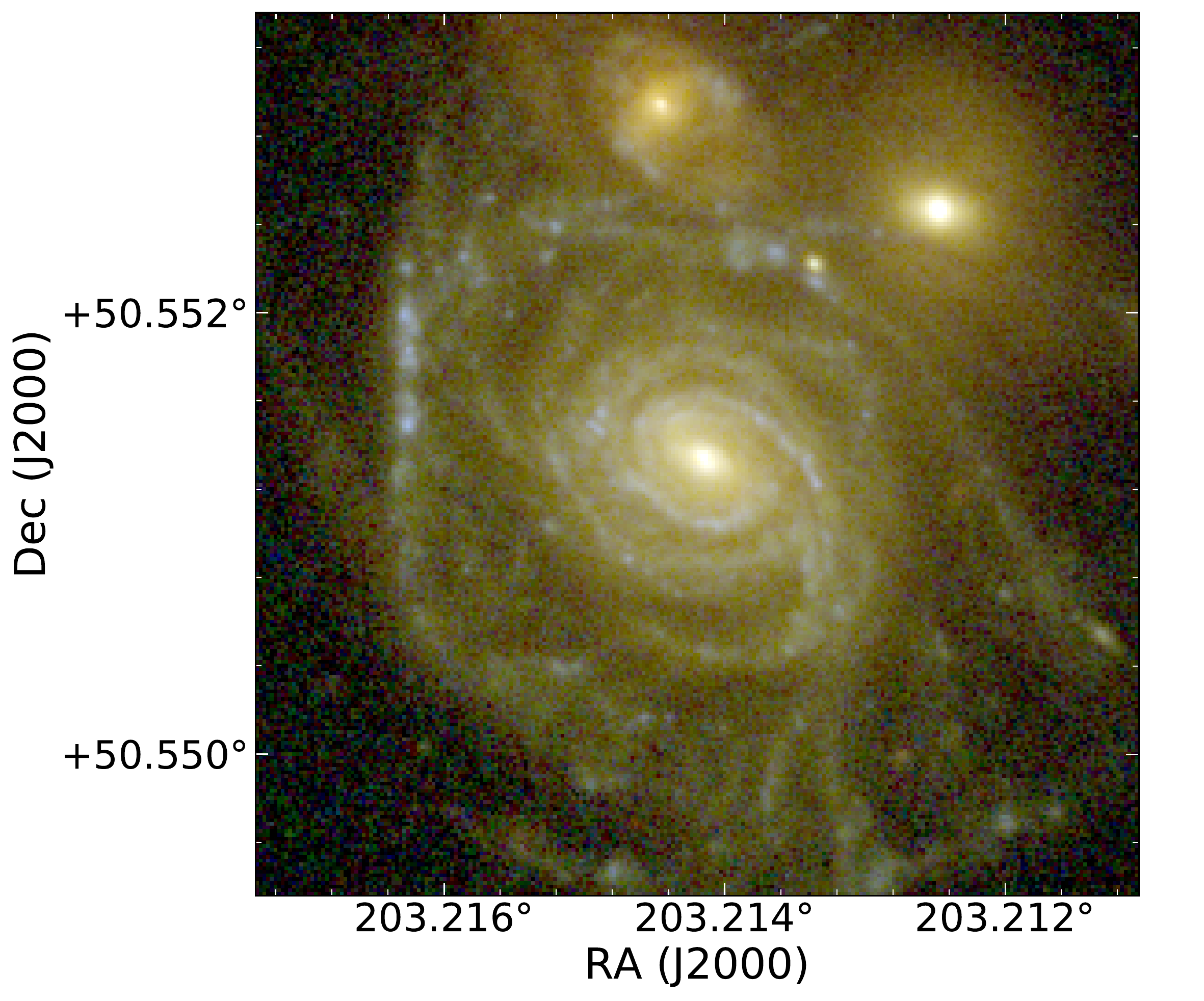}\includegraphics[width=0.62\linewidth, keepaspectratio]{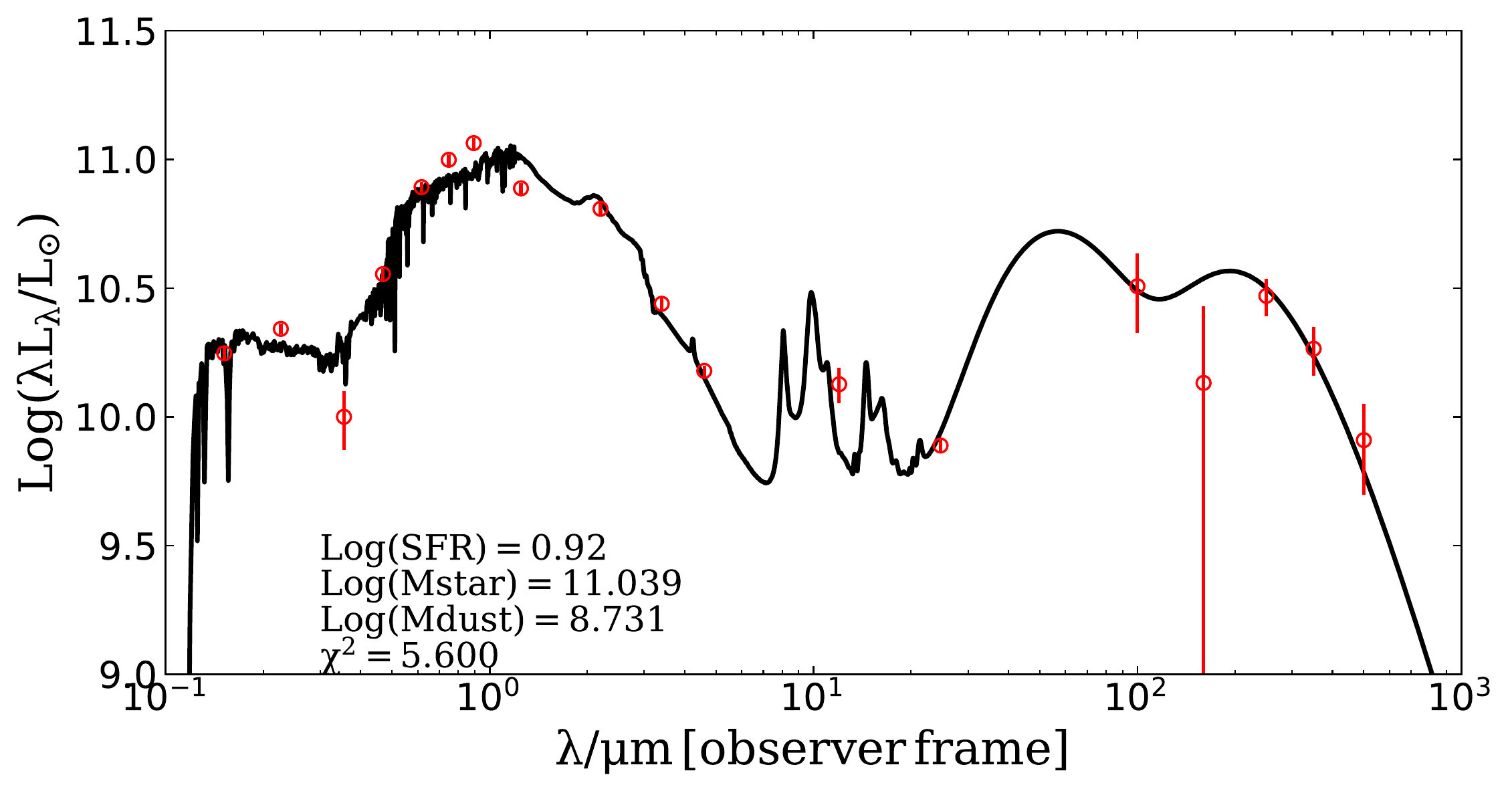}

\includegraphics[width=0.38\linewidth, keepaspectratio]{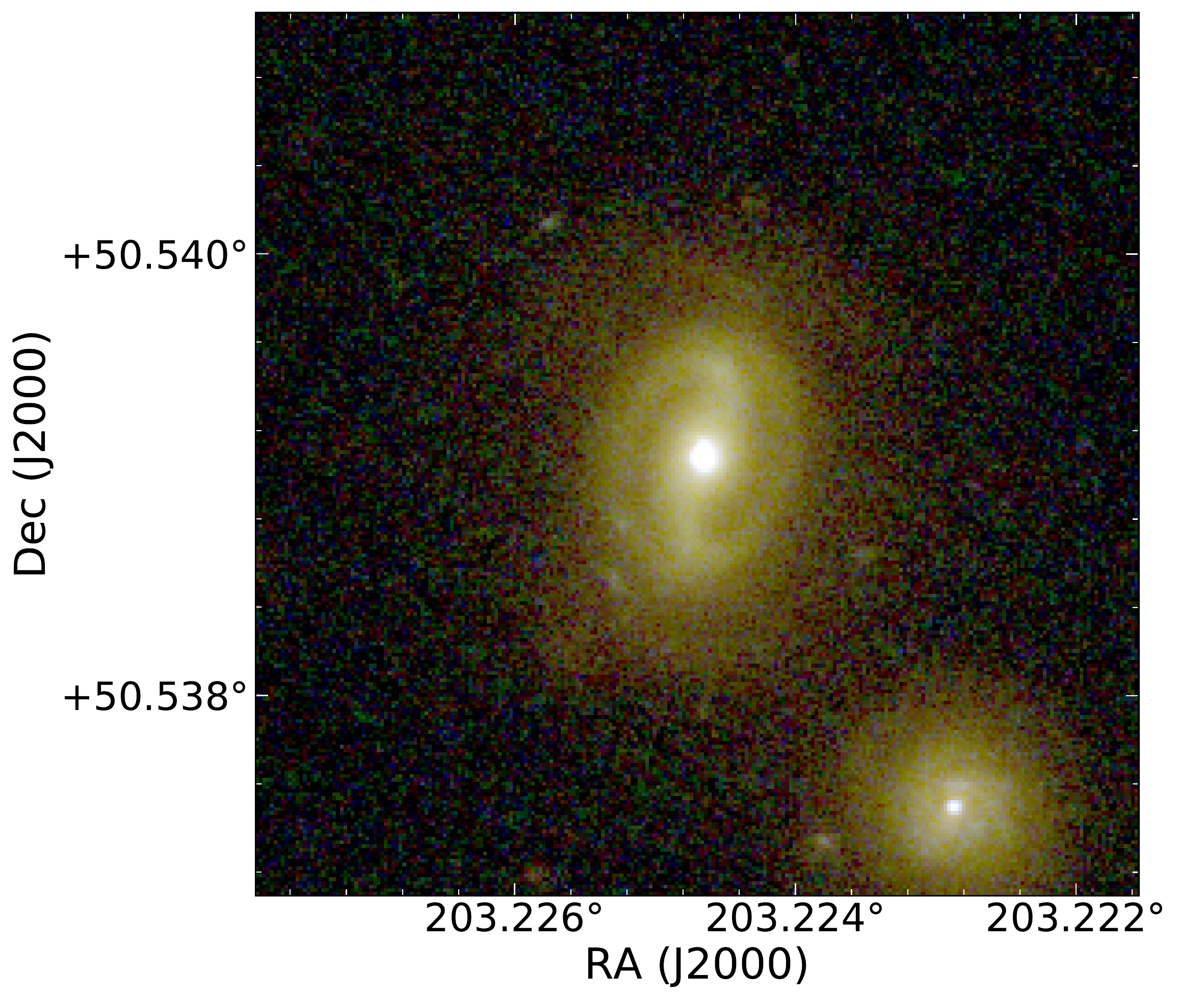}\includegraphics[width=0.62\linewidth, keepaspectratio]{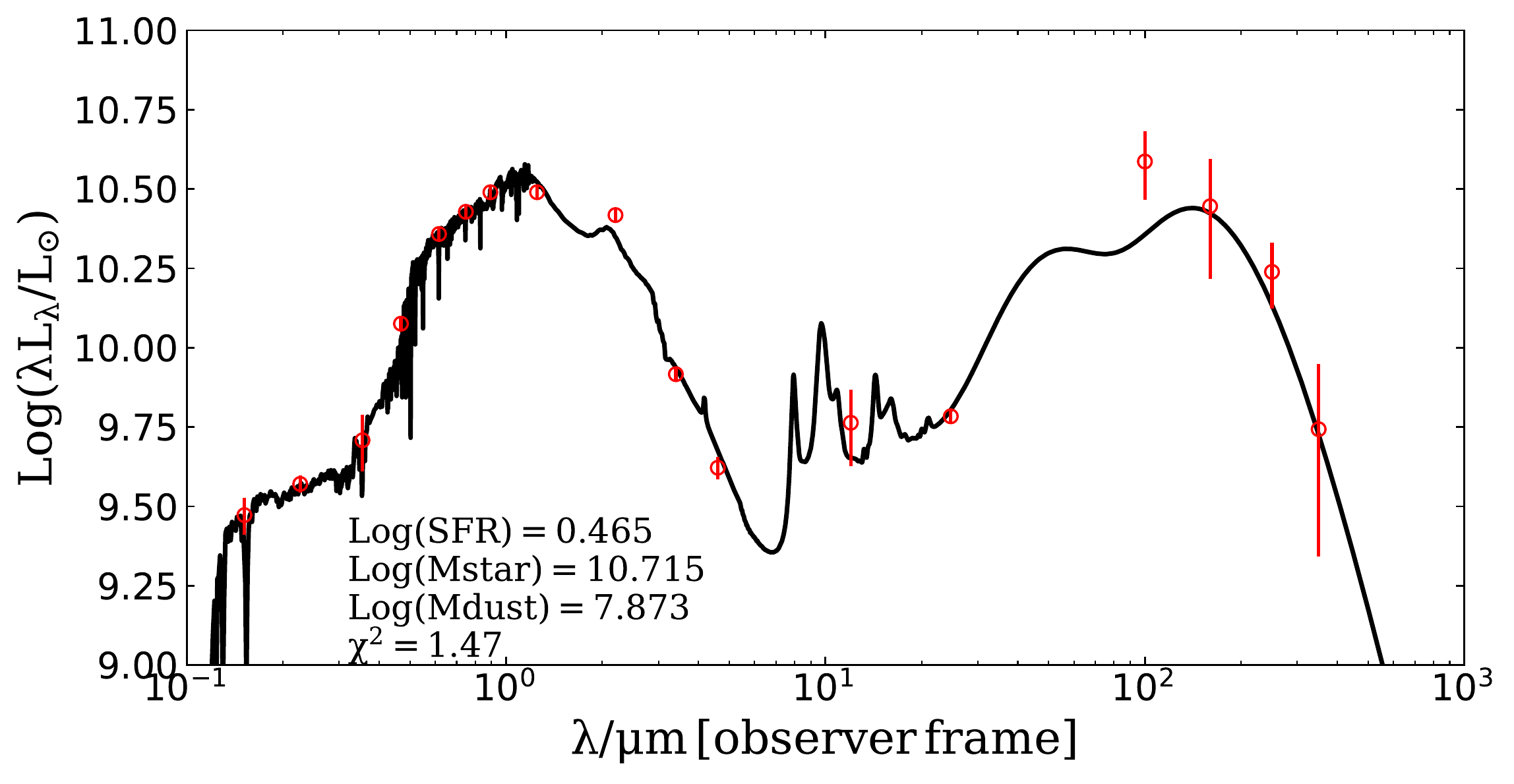}
\caption{Postage stamps and SED of two star-forming cluster members. In the left column, we see typical spiral features and colours of star-forming galaxies (30 kpc radius HST-ACS cutouts from the RELICS survey, \citealt{coe19}). In the right column, photometry (red open circles) and SED best-fit model (black curves) plots display the different properties of each galaxy and the extended wavelength coverage of the LoCuSS dataset. Salient properties obtained for each galaxy from SED fit are highlighted in text: $\rm Log (SFR [M_{\odot}\,yr^{-1}])$, $\rm Log\,(M_{dust}[M_{\odot}])$  and $\rm Log\,(M_{*}[M_{\odot}])$ and $\rm\chi^2$ of the fit. }\label{sed_image}
 \end{figure*}

Even more uncertain is what role dust plays in the environmentally driven suppression, or quenching, of star formation. Environmental processes have been shown to affect atomic gas content, resulting in truncated density profiles in the outskirts of galaxies \citep{davis13}. Environmental effects on molecular gas, and consequently on dust, are still a subject of debate (\citealt{cortese12,koyama17}), but there are measurements of the spatial distribution of dust in cluster galaxies that are consistent with it having been stripped from the galaxy (\citealt{gomez10, walter11}). Further evidence for differential dust content in clusters and the field were found in the first systematic dust surveys of the Local Universe \citep{cortese12}. While this has shown that the dust content of galaxies in clusters is different from that of galaxies in the field, the physical mechanism causing this could be any of: ram pressure stripping (\citealt{gunn72, jablonka13}), galaxy harassment \citep{moore96}, strangulation \citep{larson80} or heating from the intracluster medium (ICM) \citep{mok16}. 

Clearly, a powerful method to test models of dust formation and destruction, and their relation to star formation, is to examine how key scaling relations, such as between dust and stellar mass and dust mass and star-formation rate, vary in different environments. 
Furthermore, the cluster's dynamical state has to be taken into account. Merger events are accompanied by shock fronts, expanding through the ICM, which in turn can affect the gas and dust content in cluster galaxies.

In this paper, we concentrate on a single cluster Abell 1758 (A1758) at $\rm z=0.28$, which is known for its complex formation history forged by recent and ongoing mergers of separate clusters (Table~\ref{table_props} and Figure~\ref{wl_maps}).  A1758 is therefore an ideal laboratory in which to study the impact of local environment within clusters and cluster dynamics on dust content in member galaxies.

X-ray analysis by \citet{david04} first evidenced the absence of excess X-ray emission between A1758N and S, which originates from  merger shocks compressing the ICM. This suggested that A1758N and S have yet to interact with each other. Further analysis of \textit{Chandra} images revealed that the broadly-peaked X-ray emission to the North is associated with two prominent subclumps A1758NW and A1758NE separated by 800 kpc, and currently receding from each other, being observed some 300 Myr after the first core-passage \citep{david04}. Recently, \citet{schellenberger19} has confirmed this scenario, and additionally identified a shock front on the North side of the sub-cluster A1758NW, best-fit by a supersonic collision with Mach number 1.6, indicating a relative velocity of 2100 $\rm km \,s^{-1}$. \citet{david04} hypothesized that the South subcluster is further divided into two substructures, that will merge perpendicularly to the plane of the sky \citep{monteiro-oliveira16, schellenberger19}. The scenario of multiple clumps at different stages of merging is also corroborated by numerical simulations by \citet{durrett11} and \citet{machado15}. The zoomed-in image on A1758N with \textit{Hubble}-ACS \citep{coe19} confirms that the majority of cluster members present spheroidal/elliptical morphology. Nevertheless, disc galaxies emerge at increasing distance from the cluster cores. Distinct spiral arms, together with signatures of ram-pressure stripping \citep{ebeling19}, indicate that these galaxies are undergoing first encounter with the cluster environment. 

This paper is structured in the following manner. In Section~\ref{data}, we present the datasets used. In Section~\ref{results}, we present the methods and main results of the data analysis. In Section ~\ref{discussion}, we discuss the results and future prospects of the project. Throughout this work, we assume $\rm H_0= 70\, km \,s^{-1} Mpc^{-1}$, $\rm\Omega_M= 0.3$ 
and $\rm\Omega_{\Lambda}= 0.7$, and will not explicitly write the base (always 10) of logarithms.

\section{Observational Data}\label{data}

\begin{figure}
 \centering
 \includegraphics[width=\linewidth, keepaspectratio]{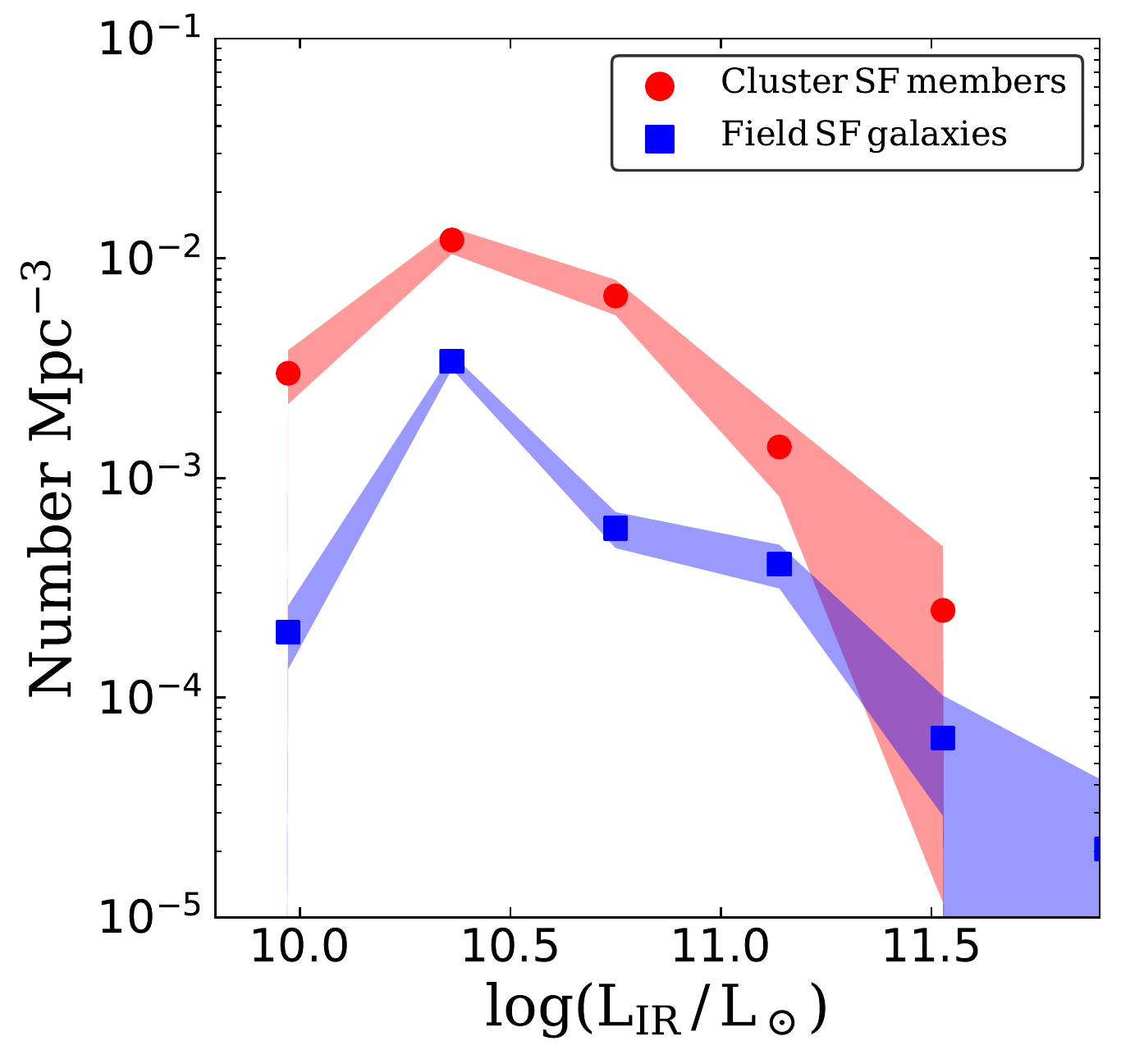}
\caption{Comoving number density of star-forming cluster members and field galaxies as a function of their total infrared luminosity. The shaded areas are $\rm\sqrt{N/Volume}$.}\label{num_flux}
 \end{figure}

A1758 is among the clusters selected for the LoCuSS survey \citep{smith10}. As a result, it benefits from the extensive coverage in both wavelength and area with GALEX (FUV and NUV), Subaru/Suprime-Cam ($g+$ and $R$ bands), UKIRT/WFCAM($J$ and $K$ band), mid-infrared 24$\rm \mu m$ with Spitzer/MIPS (reaching 90\% completeness at $400\mu$Jy) and far-infrared with Herschel (\citealt{haines10, smith10, pereira10}), both covering $\rm 25'\times 25'$ fields. In particular, as part of the LoCuSS Open Time Key Program on \textit{Herschel} A1758 was observed at 100 and 160$\rm \mu $m with PACS and at $250$, $350$, and $500\mu{\rm m}$ with SPIRE \citep{smith10}. \textit{Herschel} flux limits are 13.0, 17.0, 14.0, 18.9, 20.4 mJy from 100 to 500 $\rm \mu$m at 3$\rm\sigma$ \citep{rawle12a}. Additionally, A1758 is part of the the volume-limited high-$\rm L_{x}$ LoCuSS sub-sample of 50 clusters and has \textit{XMM}-Newton imaging (see \citealt{martino14} for further observational details). These observations were utilised to detect 39 new infalling galaxy groups surrounding 23 LoCuSS clusters, captured at their first encounter with the cluster environment \citep{haines17}. Furthermore, wide-field ($\rm \approx$1 degree diameter) optical spectroscopy with MMT/Hectospec was performed, as part of the Arizona Cluster Redshift Survey (ACReS, \citealt{haines13}).  ACReS observations provide spectroscopic redshifts for 96\% of the sources detected at $24\mu{\rm m}$ with \emph{Spitzer} down to $400\mu{\rm Jy}$.
Archival SDSS and WISE photometry was added to the data pool. In particular, we used the AllWISE Source Catalog, reaching flux limits (at SNR 5) of 54, 71, 730 $\rm \mu$Jy for 3.4, 4.6 and  12$\mu{\rm m}$ respectively \footnote{http://$\rm wise2.ipac.caltech.edu/docs/release/allwise/expsup/sec2\_1.html$}. 

In this work, we focus on star-forming galaxies, both as cluster members and field galaxies. The sample of coeval field galaxies is included as a benchmark to allow the study of environmental effects on star formation. In particular, we consider those spectroscopically-confirmed cluster member galaxies that are detected at 24$\rm \mu$m and also lie within the Herschel-PACS footprint. Field galaxies are selected from observations of 5 additional clusters from the LoCuSS survey at $\rm z<0.3$, that were observed with Herschel PACS and SPIRE instruments in the exact same way as A1758, that is covering the same sized fields ($\rm 25'\times 25'$) to the same depths in all five far-IR bands. From these data we select field star-forming galaxies  within the redshift range 0.23<z<0.30 after excluding those galaxies within $\rm 4000\, km\,s^{-1} $of the mean redshift of cluster members \citep{haines13}.  

\section{Analysis and results}\label{results}

\subsection{Number counts}

In Figure~\ref{num_count} we plot the number of spectroscopically confirmed star-forming galaxies in the cluster and field samples as a function of flux density in the $100$, $160$, and $250\mu{\rm m}$ bands, without any attempt to normalize the counts to the volume surveyed.  The number counts of star-forming field  galaxies are flatter than the number counts of star-forming clusters galaxies, especially at $100\mu{\rm m}$ and $160\mu{\rm m}$, suggesting a dearth of IR-bright galaxies in clusters relative to the field and/or differences in typical spectral shape at these wavelengths between the two samples.
Radiation at these wavelengths is characteristic of warm dust surrounding regions of star formation: spiral galaxies discs and arms are the locus of new episodes of star formation, resulting in UV emission from newly born stars which is reprocessed by surrounding dust in the IR, both of which are lacking in typical elliptical galaxies. Typical morphologies of star-forming galaxies can be seen in the left panels of Figure~\ref{sed_image} and in the Appendix (Figure~\ref{relics_cuts}) confirming the expected spiral morphology.

\subsection{Spectral energy distribution modelling}

In order to search further discrepancies between cluster and field object, we combined the entire photometric coverage to obtain the spectral energy distribution (SED) of each galaxy. The public code MAGPHYS (Multi-wavelength Analysis of Galaxy Physical Properties) by \citet{dacunha08} allows us to derive salient physical parameters of galaxies by fitting mock SED templates to the observed multi-wavelength photometric data.

For each mock template, the radiation of stars, assumed to be the only heating source, is reprocessed by dust through a two-phase grey body model. Stellar emission is computed using the population synthesis model of \citet{bruzual03} and by assuming a \citet{chabrier03} initial mass function, which is restricted to $\rm 0.1<M/M_{\odot}<100$.  The dust model includes both dense molecular clouds and a diffuse interstellar medium, following \citet{Charlot00}, re-emitting the stellar radiation through four separate components at different temperatures. In particular, these models comprise cold (15-25 K) and warm (30-60 K) dust, a hot continuum (130-250 K) from interstellar grains and polycyclic aromatic hydrocarbon emission, the total radiation of which spans between 3 and 1000 $\mu$m  (\citealt{dacunha08}, \citealt{clemens13}). Stellar population and dust models are joined to produce mock SED if they yield similar ISM luminosities, with respect to the total dust emission. The mock SED are fit to the observational data and the fit is evaluated using $\rm\chi^2$. The best-fit physical parameters are selected as the median of the probability density of the parameter values weighted by the probability $\rm exp(-\chi^2/2)$ of each fitted mock SED (\citealt{dacunha08}, \citealt{clemens13}). 

In this work, we utilise estimates of stellar mass $\rm M_{*}$, dust mass $\rm M_{dust}$, total 3-1000 $\mu$m IR luminosity $\rm L_{IR}$, the relative contributions to $\rm L_{IR}$  from birth clouds $\rm L_{BC}$, and star formation rate $\rm SFR$ (see Table~\ref{sed_prop_c} and \ref{sed_prop_f} for a list of properties of star-forming cluster and field galaxies, respectively).  The flux limits quoted in Section~\ref{data} (up to 160 $\rm \mu m$, which encloses the dust emission bump at the redshift considered here) are used as upper limits for the fluxes of the sources with no detection. This helps in constraining the models, in particular at longer wavelength, and avoiding non-physical dust masses.  Figure~\ref{num_flux} shows that the comoving number density of star-forming cluster galaxies is a factor $\sim7$ higher than the comoving number density of star-forming field members. This is not surprising given that galaxy clusters such as A1758 are overdense regions, resulting in high number counts of galaxies per unit volume, compared to the field sample. This holds true also when considering star-forming galaxies, given that the volume studied here includes the wider overdense infall region (see also \citealt{haines15}).  In literature, luminous and ultra luminous infrared galaxies (LIRG and ULIRG) are classified for having $\rm L_{IR}>10^{11} \,L_{\odot}$ and $\rm L_{IR}>10^{12} \,L_{\odot}$, respectively. It is interesting to notice the flatter trend of field star-forming galaxies above $\rm Log L_{IR} =10.5$,  suggesting a higher fraction of luminous infrared galaxies in the field than in the cluster. 

We select a final sample of cluster and field star-forming galaxies that satisfy $\rm M_{*}>10^{9}\,M_{\odot}$, $\rm L_{IR}>10^{9.8}\,L_{\odot}$ and $\rm SFR>0.1\,M_{\odot}\,yr^{-1}$.  The two samples comprise 90 and 68 cluster and field star-forming galaxies, respectively. This selection extends beyond the level at which the samples can be considered complete; see for example the lowest luminosity bin in Figure~\ref{num_flux}.  The results described in the following sections are insensitive to whether we restrict our samples to be more complete.  We interpret this as indicating that any effects of incompleteness at the faint limit of our samples affect both samples in the same manner.

\subsection{Dust mass, stellar mass, and SFR distributions}\label{allfield}

\begin{figure*}
 \centerline{
  \hspace{-0.3cm}
  \includegraphics[width=0.5\linewidth]{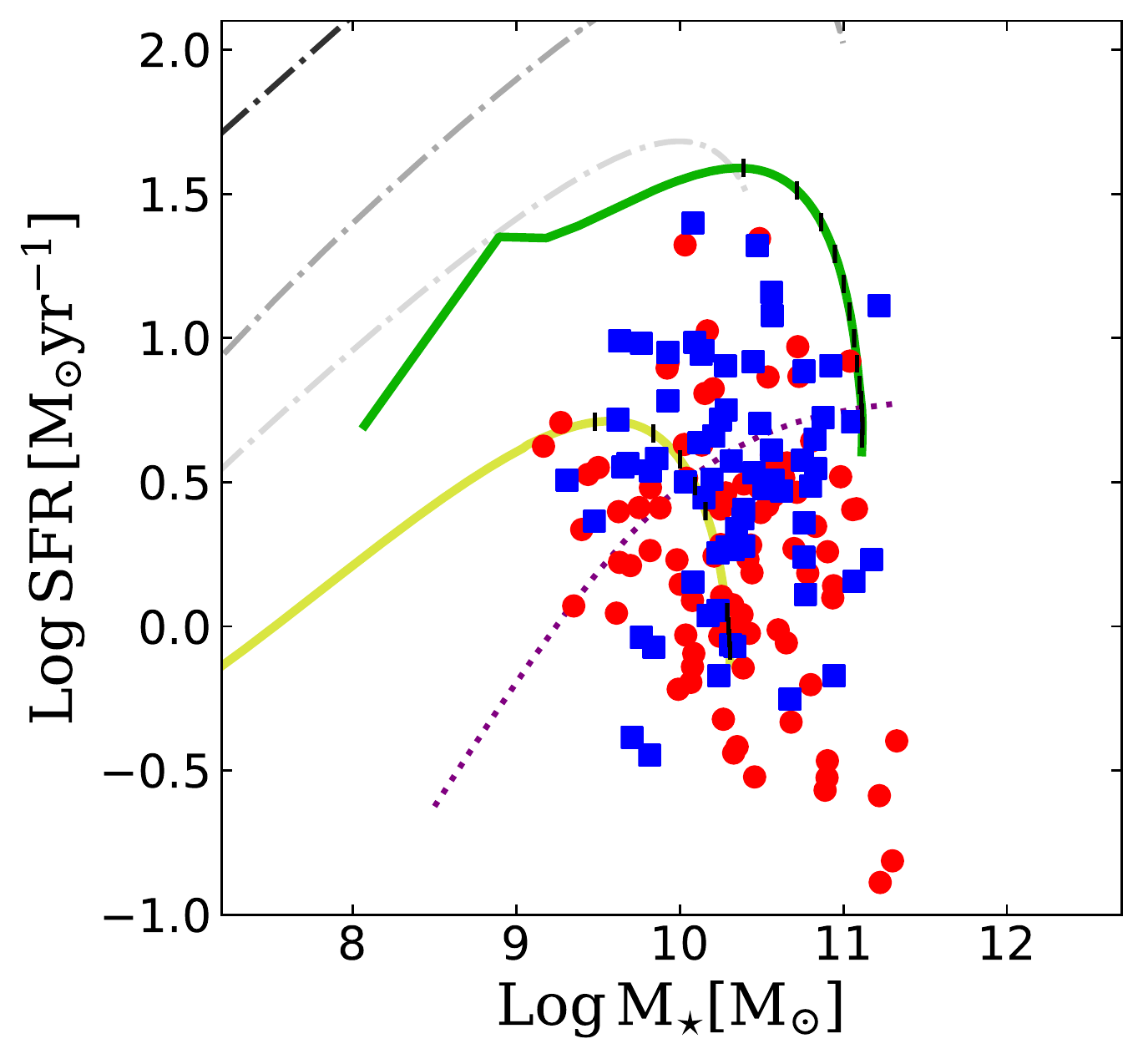}
  \hspace{2mm}
  \includegraphics[width=0.475\linewidth]{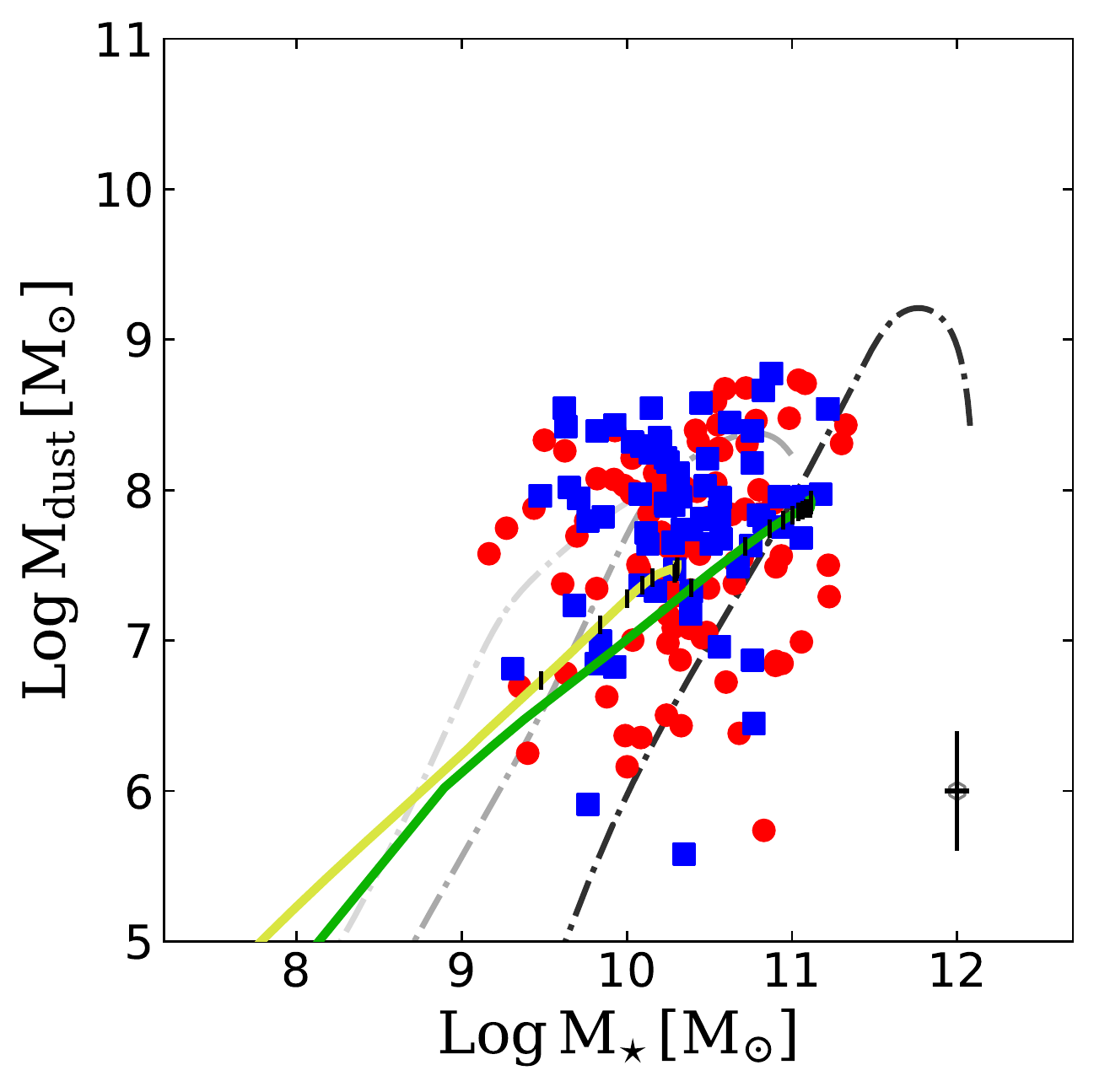}
 } 
 \centerline{
   \includegraphics[width=0.6\linewidth]{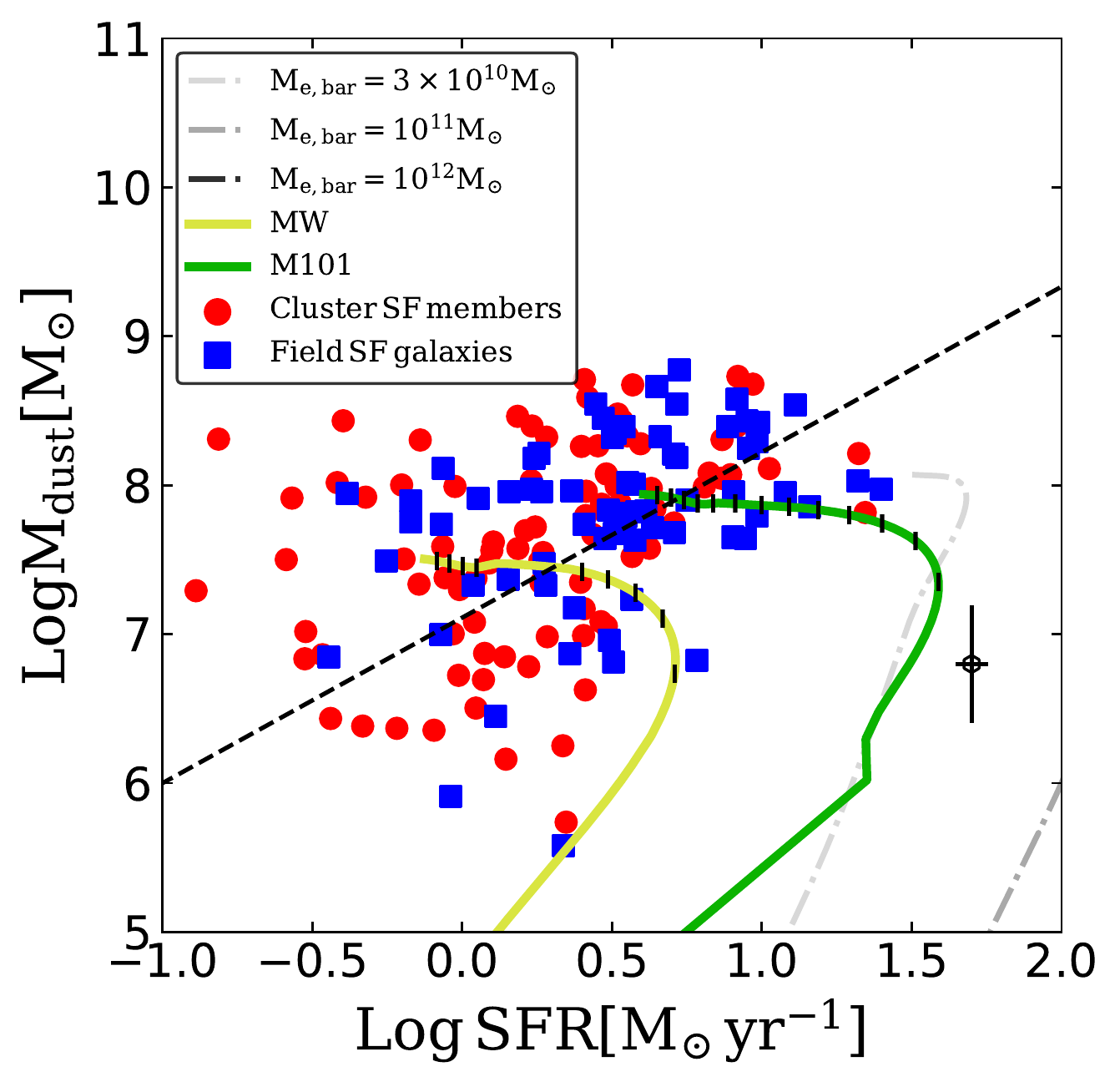}
   }
\caption{Top left: stellar mass plotted against SFR for cluster and field star-forming galaxies marked as red circles and blue squares, respectively. The dotted purple line marks the star-formation main sequence at $\rm z\approx0.35$ from the COSMOS survey \citep{lee15}. Top right: stellar versus dust mass relation. Bottom panel: dust mass plotted versus SFR for cluster members and field star-forming galaxies. The dashed line shows the fit to a sample of star-forming galaxies with $\rm z<0.22$ from \citet{dacunha10}. In each panel, the error bars represents the dispersion associated to the median value of the plotted quantities, and connects the 16th to the 84th percentiles of each parameter distribution. Overplotted are the theoretical tracks computed in \citet{calura16}, marking the evolution of stellar and dust mass and star-formation rate according to different galaxy evolution recipes. In particular, dashed lines and continuous lines correspond to spheroidal starburst galaxies ($\rm M_{e, bar}$) and spirals  of increasing baryonic mass, respectively (see Section~\ref{allfield}). Spirals are modelled in order to reproduce Milky Way (MW) and M101- type galaxies. }\label{sfr_mdust}
 \end{figure*}

Dissecting stellar and dust masses, and SFR aims at measuring the impact of infall onto cluster on the star formation cycle and how it echoes among these properties. This, when compared to the coeval reference field sample, helps in constraining quenching since dust and stars production/destruction cycle is susceptible to processes acting in different environments.

\begin{figure*}
 \centering
\includegraphics[width=\linewidth, keepaspectratio]{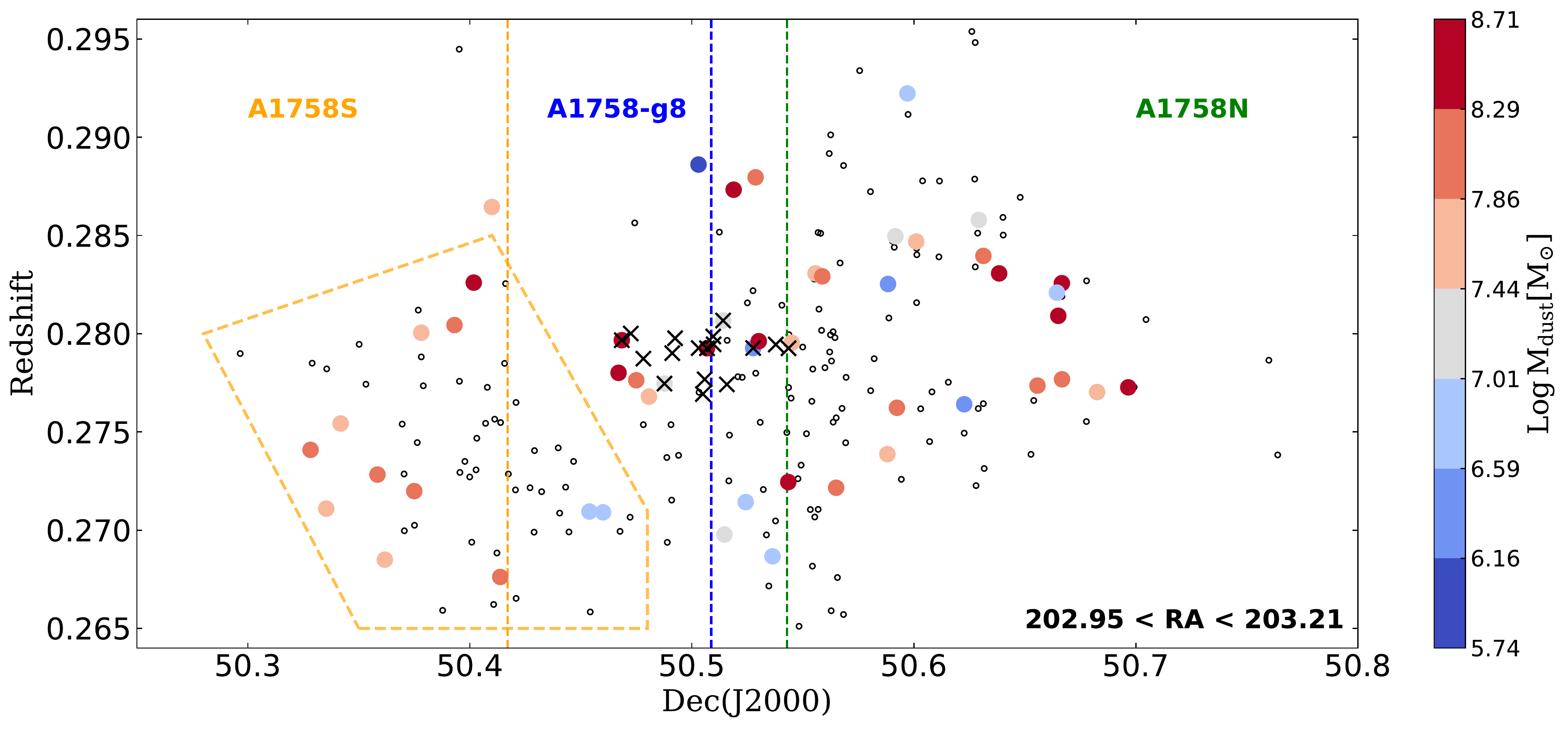}
\caption{Declination plotted with respect to redshift for the spectroscopically confirmed members (black points) of the cluster A1758, divided among the two main subsystems A1758N and A1758S (enclosed in the orange dashed polygon), and A1758-g8 (black crosses). Vertical dashed lines mark the declination of the X-ray centres of each subsystem. The colour-map codes dust mass of star-forming members in logarithmic scale.}\label{dec_z}
 \end{figure*}

Overall, both cluster and field star-forming galaxies span similar values of stellar and dust mass and SFR, as it can be seen in the different panels of Figure~\ref{sfr_mdust}.  As highlighted by the average errors distributions in each panel of Figure~\ref{sfr_mdust}, dust masses have higher fractional errors ($\rm \approx 40\%$) with respect to stellar masses and SFR ($\rm \approx 20\%$). Despite the high values, such uncertainties are typical for dust mass estimation. In the top-left panel of Figure~\ref{sfr_mdust}, we can see that our sample presents overall comparable values of stellar mass and star formation rates to the COSMOS low redshift sample ($\rm z \approx 0.35$, \citealt{lee15}). A clear correlation can be seen in the bottom panel of Figure~\ref{sfr_mdust} between SFR and $\rm M_{dust}$ for cluster and field galaxies. This is not surprising given that dust emission consists of re-processed light from newly formed stars. Our sample shows consistent values of star-formation rate and dust masses with respect to the star-forming galaxies with z < 0.22 from \citep{dacunha10}. Interestingly, despite the scatter, field star-forming galaxies show on average higher values of SFR compared to cluster members across the probed stellar mass range, and in particular towards the high mass end.  In particular, galaxies with stellar masses of $\rm M_\star>10^{10}M_\odot$, the mean SFR per unit stellar mass of field and cluster star-forming galaxies is $\rm 2.07 \times 10^{-10} yr^{-1}$ and $\rm 1.28 \times 10^{-10} yr^{-1}$  respectively, at $\rm 1.5\sigma$. This is consistent with the discrepancy in specific SFR between cluster and field star-forming galaxies found in \citet{haines13}, suggesting slow quenching (occurring on timescales between 0.7-2.0 Gyr) of star formation in galaxies upon accretion on clusters.

Despite the difference between the typical SFR of field and cluster star forming galaxies, both populations are more closely described by evolutionary models of spiral galaxies than by proto-spheroidal starburst galaxies. This can be seen in the better agreement of the Milky Way- and M101-like models with our data in Figure~\ref{sfr_mdust} than the starburst galaxy models from \citet{calura16}. These models describe galaxies spanning increasing range of masses. In particular, spheroidal galaxy models assume total baryonic masses of $\rm 3\times 10^{10}$, $\rm 10^{11}$ and $\rm 10^{12}\,M_{\odot}$.  Spiral-like models span approximately 13 Gyr of evolution, while spheroidals are limited to approximately 1 Gyr. For this reason, we display time-steps of 1 Gyr only for spiral-like models. Each model evolution starts at low values of stellar mass and SFR in the top and bottom panels, respectively. This comparison helps highlighting the different evolutionary path between star-forming field galaxies and cluster members. In particular, we can notice the absence of recent starbursting galaxies in our sample, which are much better described by long-lasting star formation history of typical spiral galaxies. In addition, we confirm that dust and stellar masses scale almost indistinguishably, even for vastly different object, and that SFR is a key parameter to better constrain the evolution of these galaxies.

\subsection{Spatial and phase space distribution of star-forming cluster galaxies}

We have already seen in the skymaps shown Figure~\ref{wl_maps} that star-forming cluster galaxies appear to reside preferentially in the cluster outskirts.  We now investigate this further using the redshifts from ACReS to disentangle the memberships of A1758N, A1758S, and A1758-g8.  Figure~\ref{dec_z} shows the distribution of cluster members in declination versus redshift plane. This display helps to highlight the different components of A1758. To facilitate identification, the declination of the main cluster components, A1758N and S together with A1758-g8, are marked by vertical dashed lines. We can clearly notice a gap in the galaxy distributions belonging to the two clusters A1758N and A1758S, confirming that the two systems have yet to encounter each other and mix their galaxy populations. Furthermore, star forming galaxies are funneled towards A1758N and S not isotropically, but rather along separate accretion paths. Overall, this suggests that A1758N and S belong to two separate virialized dark matter halos, which are not yet connected.  

Colour map codes the dust mass of star-forming cluster members. The dearth of star-forming galaxies is evident in the core of both subclusters, together with the increase of dusty galaxies towards the cluster outskirts, which correspond to the outer edge of merger shocks presented in \citet{schellenberger19} (see also X-ray countour edge in Figure~\ref{wl_maps}). North of A1758, an elongation in galaxy distribution suggests channels of accretion, potentially filamentary structures, which are not traced by the X-ray emission. Similarly, group members are located perpendicular to the axis connecting A1758N and S, and might suggest a further infall channel toward the northern clump. Star-forming cluster members do not follow the distribution of virialized passive cluster members, as confirmed in Figure~\ref{r_vel_dust}. In this plot, both clusters A1758N and A1758S, and group members are stacked according to their projected distances from their respective cluster/group centre and  velocities with respect to their average halo redshift. In the case of A1758N, the centre is fixed to be half-way between the two northern BCGs, and for A1758S and  A1758-g8 the centre corresponds to the coordinates of the southern BCG and X-ray emission peak, respectively. Projected distances and  velocities are further scaled by their halo $\rm r_{200}$ and velocity dispersion $\rm\sigma$, respectively. When compared to the rest of cluster members, star-forming galaxies present a flat distribution of velocities. Velocities displaying a Gaussian distribution peaked around low values are typical of the old virialized population of cluster members. On the other hand, a flatter distribution of  velocities is characteristic of infalling objects. The kurtosis of the  velocity distribution of star-forming galaxies, which include both clusters and group members, is $\rm \gamma = - 0.83 \pm 0.22$. which is inconsistent at $\rm 3.7 \sigma$ with the $\rm \gamma = 0.0$ of a Gaussian distribution.  Jointly, this shows that star-forming galaxies have been recently accreted onto the cluster potential \citep{haines15}.

\subsection{Dust-to-stellar mass ratio}

\begin{figure}
 \centering
\includegraphics[width=0.8\linewidth, keepaspectratio]{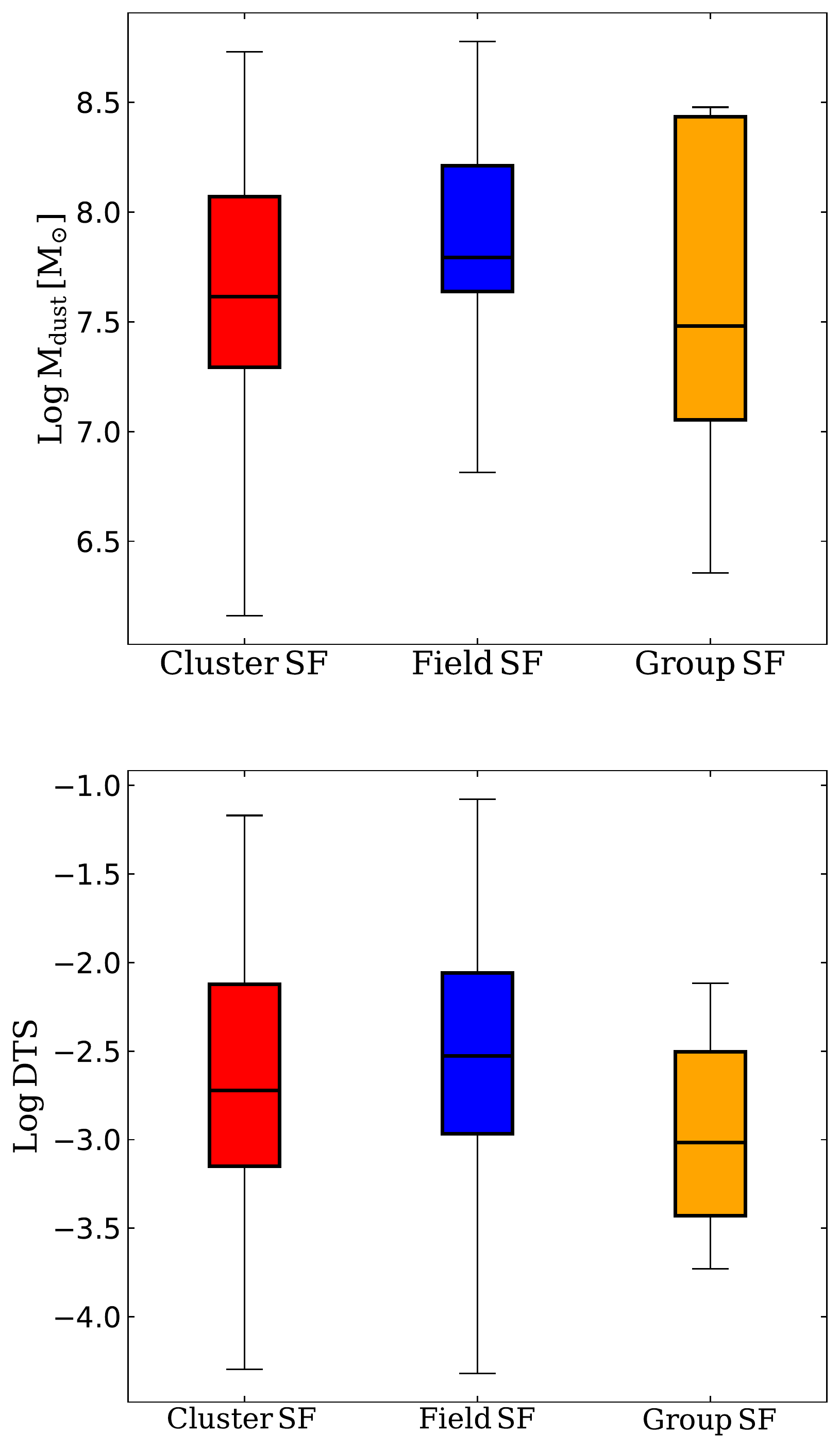}
\caption{$\rm M_{dust}$ and DTS for cluster, field and group star-forming galaxies. The solid horizontal line shows the mean of each sample, the coloured box encloses the upper and lower quartiles, and whiskers extend to an additional $\rm 50\%$ of interquartile range, enclosing approximately 3$\rm\sigma$.}
\label{box_dust}
 \end{figure}

\begin{figure*}
 \centering
  \includegraphics[width=0.45\linewidth, keepaspectratio]{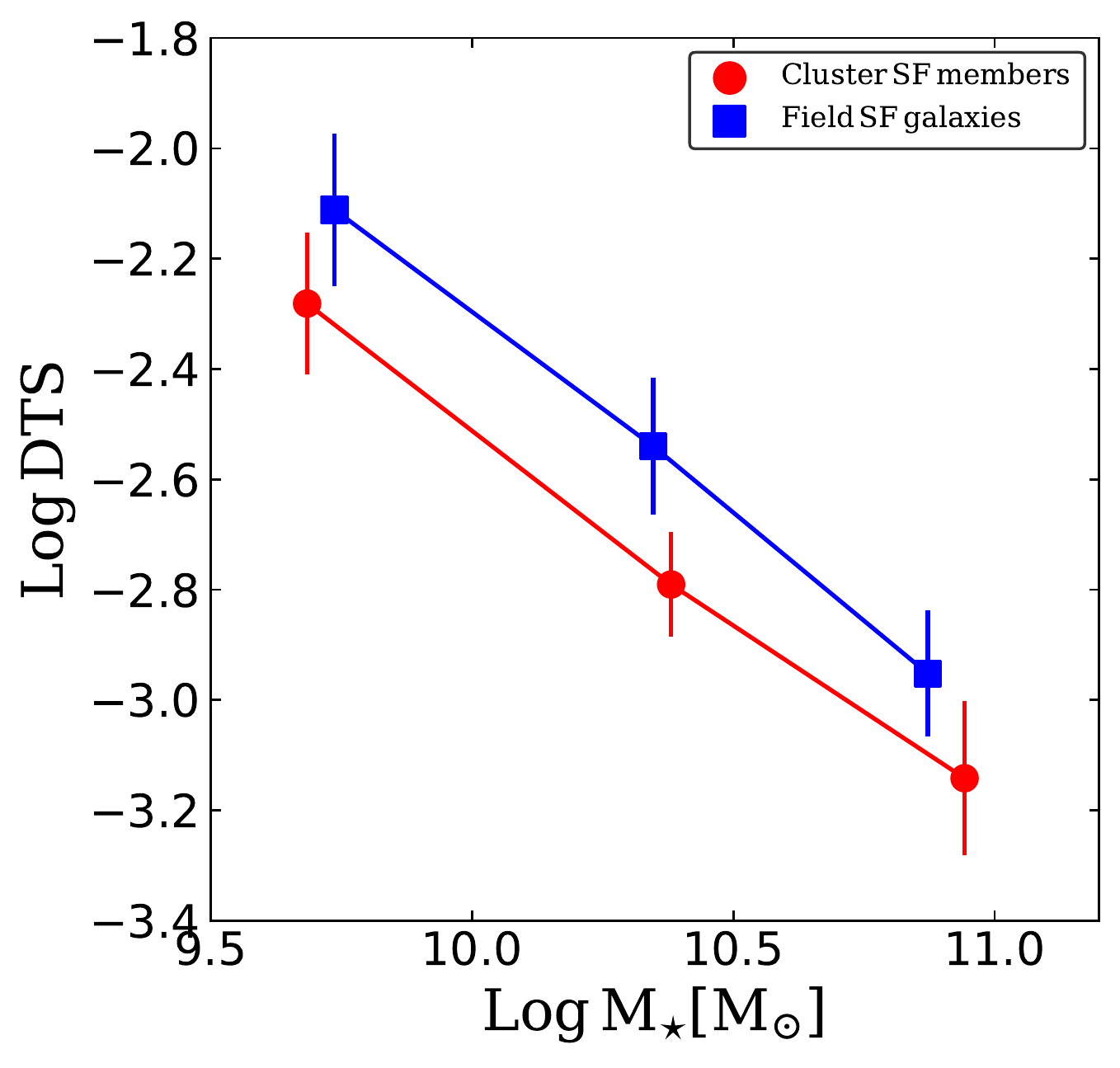}
  \includegraphics[width=0.465\linewidth, keepaspectratio]{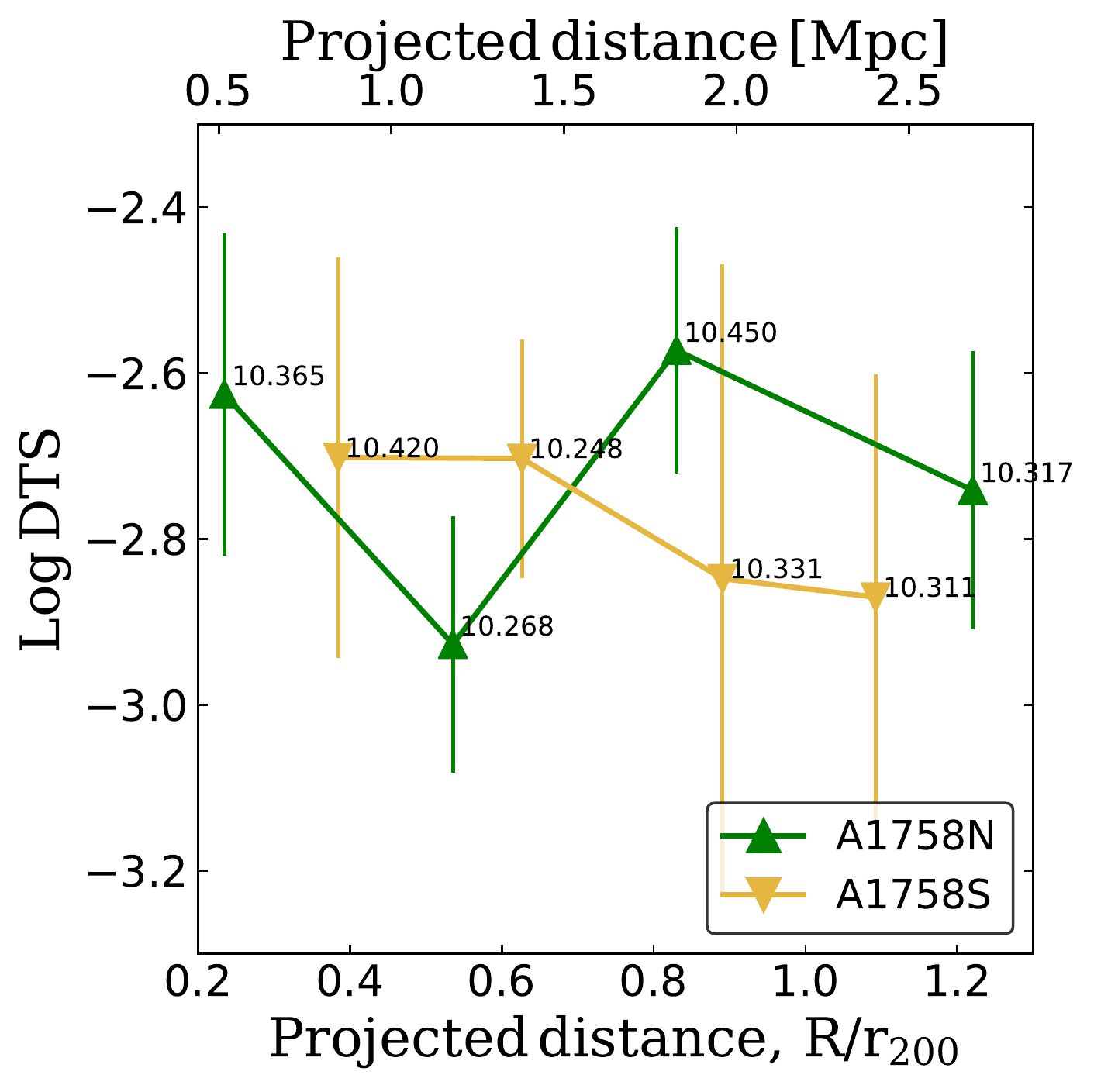}
\caption{Left panel: DTS mass ratio in equi-numeric bins plotted with respect to stellar mass. Red circles and blue squares mark field and cluster star-forming galaxies respectively.  Right panel: 1-d profile of the DTS of the stacked star-forming cluster members of both A1758 N and S with respect to projected clustercentric distance in units of $\rm r_{200}$. The top-axis is computed using a mean $\rm r_{200} = 2.2 \,Mpc$. The green and yellow profiles are computed with respect to the centre of A1758N and A1758S, respectively. Each point corresponds to the mean quantity per distance bin, each of which contains on average $\approx$ 20 and $\approx$ 9 galaxies for A1758N and S, respectively.  Error bars show the 1$\sigma$ confidence interval. The mean Log $\rm M_{\star}$ in each projected distance bin is overplotted. }\label{dust_sf}
 \end{figure*}

A useful quantity to assess the efficiency of the dust duty-cycle in galaxies is the dust-to-stellar (DTS) mass ratio. This quantity accounts for the amount of dust per unit stellar mass contained in each galaxy.  We present here the first study of the DTS ratio for a sample of cluster and field star-forming galaxies at intermediate redshift. We can see the global values of $\rm M_{dust}$ and $\rm DTS$ for cluster, field and group star-forming galaxies in Figure~\ref{box_dust}. As previously mentioned, star-forming galaxies in both cluster and field present comparable properties, within $\rm \approx 3\sigma$. More interestingly, subdividing further both samples in bins of stellar mass helps in highlighting a discrepancy between them. Left panel of Figure~\ref{dust_sf} shows the DTS ratio as a function of $\rm M_{*}$. We see a linear trend of the DTS with respect to stellar mass, for both cluster and field galaxies. More massive galaxies have less fierce star formation with respect to smaller ones, per unit stellar mass. In addition, cluster galaxies present lower values of dust per unit stellar mass, suggesting further consumption/destruction due to environment, with respect to field galaxies. Overall, we measure a shift towards lower values of the DTS in cluster galaxies, with respect to field objects,  at $\rm \approx \,2.4 \sigma$, when adding together the significance in each bin. Interestingly, the displacement between DTS in field and cluster galaxies parallels that seen in \citet{haines13} when comparing specific-SFR of cluster and field star-forming galaxies. 

\begin{figure*}
 \centering
\includegraphics[width=0.85\linewidth, keepaspectratio]{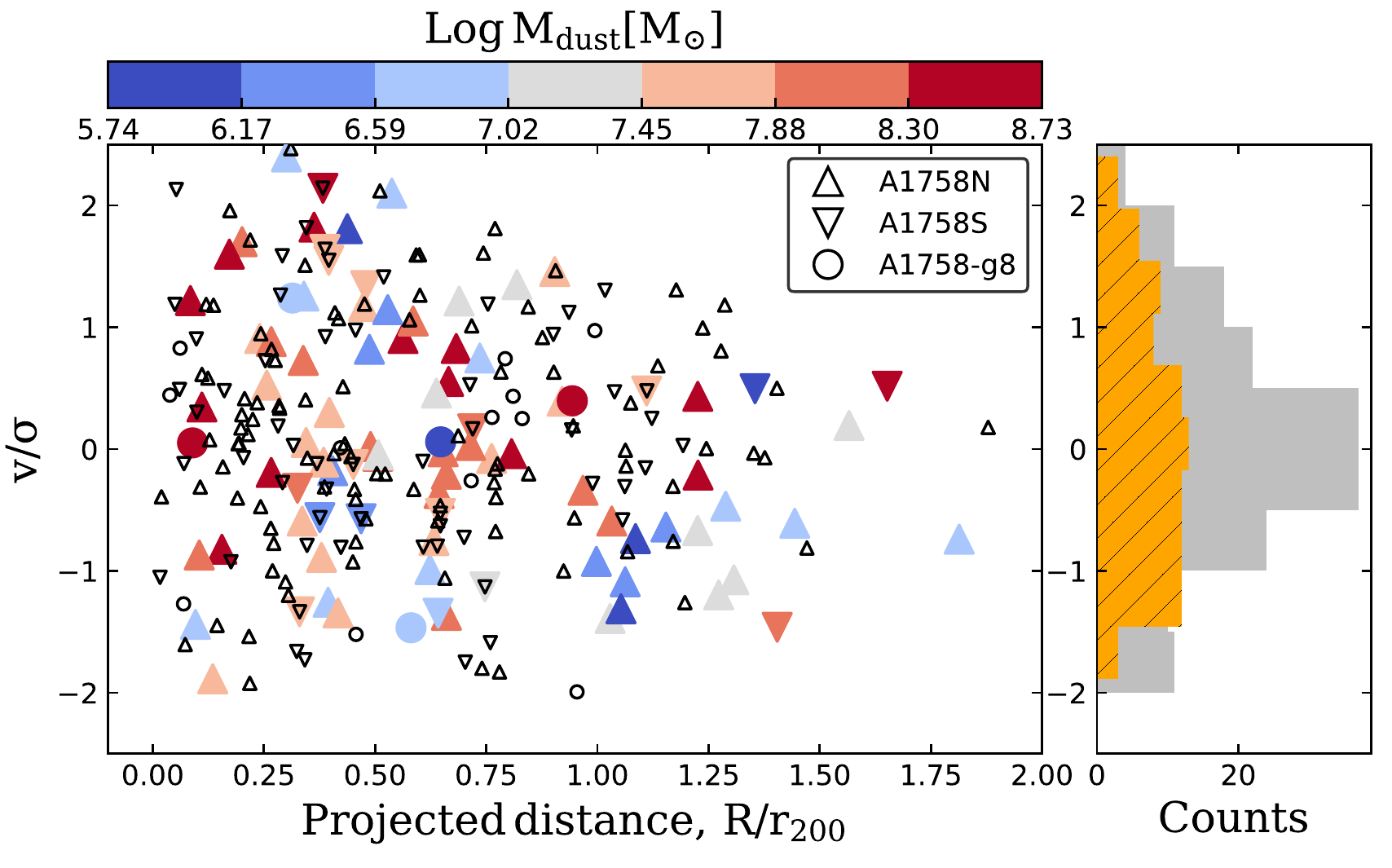}
\caption{Left panel: projected distance from the halo centre plotted against  velocity for the stacked sample of subclusters A1758N (up-triangles), A1758S (down-triangles) and A1758-g8 members (circles), in units of $\rm r_{200}$ and velocity dispersion of their parent halo. The colour-bar codes dust masses $\rm Log{M_{dust}}$. Empty black symbols mark the non star-forming spectroscopically confirmed cluster members. Right panel: distribution of the velocities of star-forming and non star-forming cluster members plotted as orange hatched and grey histograms, respectively.}
\label{r_vel_dust}
 \end{figure*}

Right panel of Figure~\ref{dust_sf} shows the DTS of the the stacked population of cluster star-forming members according to their projected distance from both A1758N and A1758S centres, as computed for Figure~\ref{r_vel_dust}. This plot shows that among star-forming cluster members there is no strong radial trend of the DTS ratio. A dip in the DTS profile can be seen around $\rm \approx 0.5 R/r_{200}$ with respect to the northern centre. This distance corresponds to the location of  A1758-g8 (see Figure~\ref{dec_z}), in the case of A1758N. We notice that the DTS value for star-forming cluster members in this bin would be further decreased with the inclusion of group star-forming members, suggesting that group galaxies have lower values of DTS, compared to the average population of star-forming cluster members at that clustercentric distance. We also report that no significant mass segregation is evident among star-forming cluster members, as shown by the average stellar masses in each bin in the right panel of Figure~\ref{dust_sf}. 

\subsection{Spectral shape: cluster versus field} 

\begin{figure*}
 \centering
\includegraphics[width=0.55\linewidth, keepaspectratio]{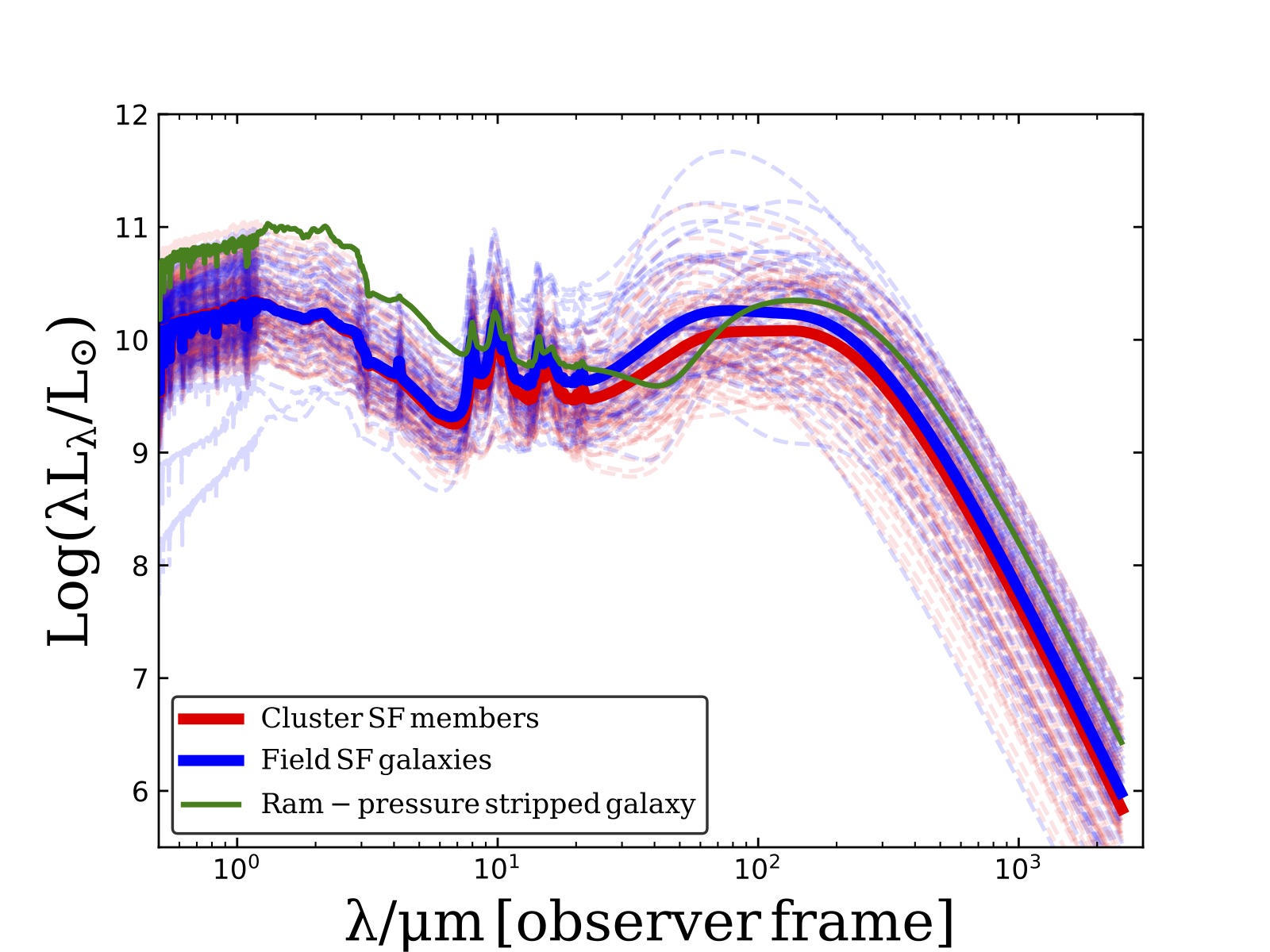}\includegraphics[width=0.4\linewidth, keepaspectratio]{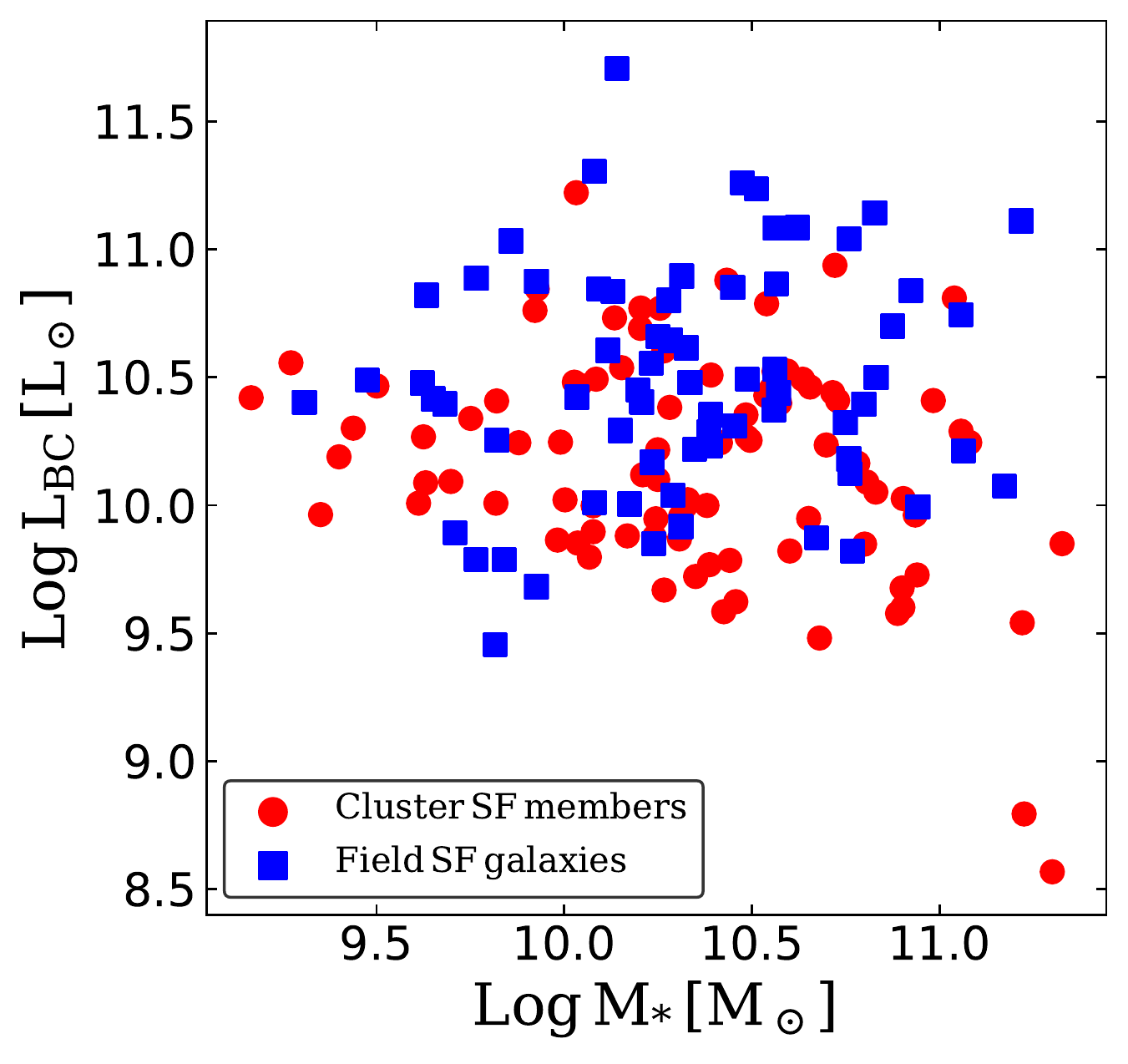}
\caption{Left panel: individual (dashed lines) and mean (solid line) SED for both clusters (red) and field (blue) star-forming galaxies. Overlaid in green, the SED of the ram-pressure stripped galaxy described in \citet{ebeling19}. Right panel: infrared luminosity from birth clouds $\rm L_{BC}$ versus stellar mass of both clusters (red circles) and field (blue squares) star-forming galaxies, which is emitted preferentially around 100$\rm\mu$m.}
\label{evo_sed}
 \end{figure*}

Notwithstanding the large scatter among the cluster and field star-forming galaxies, the mean SED of each galaxy sample shows a clear difference at $\sim100\mu{\rm m}$, with cluster galaxies being fainter than field galaxies (left panel of Figure~\ref{evo_sed}). Emission at these wavelengths is dominated by warm dust, polycyclic aromatic hydrocarbides and mid-infrared continuum, which are preferentially located in the surroundings of birth clouds \citep{dacunha08}.  Therefore emission from these components appears reduced in cluster star-forming galaxies with respect to coeval field objects.  Furthermore, we overplot our SED of the ram-pressure stripped galaxy from \citet{ebeling19}. The dip in the SED of this galaxy at $\sim50\mu{\rm m}$ shows that the emission from birth clouds is reduced in this galaxy, suggesting ram-pressure stripping as a channel for dust removal/destruction. This is further supported by the comparison of IR luminosity from stellar birth clouds $\rm L_{\rm BC}$ (dominated by the diffuse warm dust component) with stellar mass for cluster and field galaxies in Figure~\ref{evo_sed}. At a fixed stellar mass, field star-forming galaxies have higher $\rm L_{\rm BC}$ than cluster star-forming galaxies.


\section{Summary and discussion}\label{discussion}

We aim to progress the current knowledge on star-formation quenching  by studying how dust in star-forming galaxies is affected by the local environment within clusters.  In particular, this work is the first to explore dust evolution in a sample of star-forming galaxies in the dynamically complex cluster A1758 and across its main substructures.  A1758 allows us to assess the effect of  merger-induced shocks on star-formation and dust properties of clusters members.  Uniquely, the LoCuSS dataset used here allows for a direct comparison between star-forming cluster members with a coeval sample of mass-matched star-forming field galaxies. Specifically for A1758, our data span far-UV to far-IR wavelengths and include imaging with \emph{GALEX}, Subaru, SDSS, UKIRT, \emph{WISE}, \emph{Spitzer}, and \emph{Herschel}, plus 96\% complete spectroscopic follow-up of candidate star-forming galaxies from ACReS observations with Hectospec on MMT.  Galaxies considered here have been selected for having $\rm M_{star}>10^9 \, M_{\odot}$, $\rm SFR > 0.1\,M_{\odot}\,yr^{-1}$.

We find that star-forming galaxies, whether located in clusters or in field, span a similar range of stellar and dust masses, whilst the IR luminosity and SFR of star-forming cluster galaxies are lower than star-forming field  galaxies.  The dust-to-stellar mass ratio (DTS) of cluster star-forming galaxies is a factor $\approx 32\%$  lower than that of field star-forming galaxies at $\rm 2.4\sigma$ significance. This result implies an effect of the cluster environment on both the dust content, and the SFR of star-forming galaxies.  Among cluster members, DTS appears to vary little with respect to clustercentric distance in both the north and south part of the cluster, with the exception of galaxies within an infalling X-ray group which overall have lower values of dust per unit stellar mass.

Galaxy members of A1758N and S are distributed separately in position and redshift space, suggesting that the two subclusters belong to separate virialized dark matter halos. This is further confirmed by analysing the distribution of star-forming galaxies. Star-forming members are distributed towards the cluster outskirts, distant from the actively merging cores of A1758N, A1758S and the merger shocks. These galaxies are being accreted along separate accretion paths, rather than isotropically. We verified the absence of additional virialized substructures associated to star-forming cluster member, via a comparison with diffuse X-ray emission and weak lensing mass maps from \citet{haines17} and  \citet{okabe16}, respectively. This suggests that these galaxies are infalling in isolation. Isolated galaxies do not suffer from pre-processing in groups prior to their infall, thus preserving field-like properties. The combination of position and  velocity classifies the majority of cluster star-forming members as recent or back-splash infallers, i.e. approaching or receding from their first encounter with the cluster core \citep{haines15}. These galaxies have spent limited time in the harsh ICM, whereas galaxies infalling within the group have already being processed by the hot intragroup medium. This has been proven effective in reducing the SFR \citep{bianconi18}, and appears to similarly influence the dust content of galaxies.

Our study extends to higher redshift recent findings on dust consumption and destruction from observational campaigns on local clusters \citep{cortese10}, which showed evidences of truncation in the radial distribution of multi-phase gas and dust harboured in discs of star-forming galaxies. This suggests outside-in removal processes due to the cluster environment (\citealt{rawle12b,finn18}). Overall, the bulk of dust mass, which is locked in cold clumps preferentially within the plane of galaxies, appears unaffected by the cluster environment on the timescales considered here.  We observe primarily the decrease emission from warm dust grains in cluster galaxies with respect to field objects. This reflects the reduction of ionizing radiation from newly born stars, following the decrease of SFR, but also has been associated to the destruction of small, not shielded, dust grains due to ICM emission (\citealt{bocchio14, gjergo18}).  Sputtering is proposed to destroy dust grains on very short ($\rm 10^4\,yr$) timescales. For sputtering to be effective, dust has to be removed from the galactic plane where it is shielded from the surrounding ICM radiation. Dust in cluster galaxies is prone to being removed or destroyed by ram-pressure or shocks passing through the galaxy, resulting in the dearth of infrared luminous galaxies in the cluster cores.  Hence, the timescale of ram-pressure stripping ($\rm > 10^9 \,yr$) is dominating sputtering of dust particles. Interestingly, detailed thermal conduction formalism has been recently included in the framework of cosmological simulations, showing its efficiency in distributing the energy from the hot ICM into galaxies and reducing their star formation \citep{kannan17}.

We present here a systematic panchromatic study of spectroscopically confirmed star-forming cluster galaxies at intermediate redshift, which allows us to explore in detail the connection between local environment, star formation and dust. In this work, we conclude that A1758 is confirmed to be an actively evolving cluster, composed of two main subsystems North and South which belong to dynamically separated large-scale structures, and are accreting galaxies along separate paths, rather than isotropically. We observe the effect of the local cluster environment echoing among the properties of star-forming members, which include morphology, star formation rate, dust emission and masses. In particular, star-forming cluster members present diminished values of star formation and dust masses with respect to a coeval sample of star-forming field galaxies. We measure directly a decrease in the emission of birth clouds  in star-forming cluster members, with respect to field galaxies. This suggests that the timescale for transformation upon accretion to the cluster is slow. Among the proposed mechanisms responsible for the transformation, we include ram-pressure stripping, harassment, strangulation, heat conduction from the ICM and ICM shocks removing/destroying dust in recently accreted galaxies.

\section*{Acknowledgements}
\begin{figure*}
\centering
\includegraphics[width=0.32\linewidth, keepaspectratio]{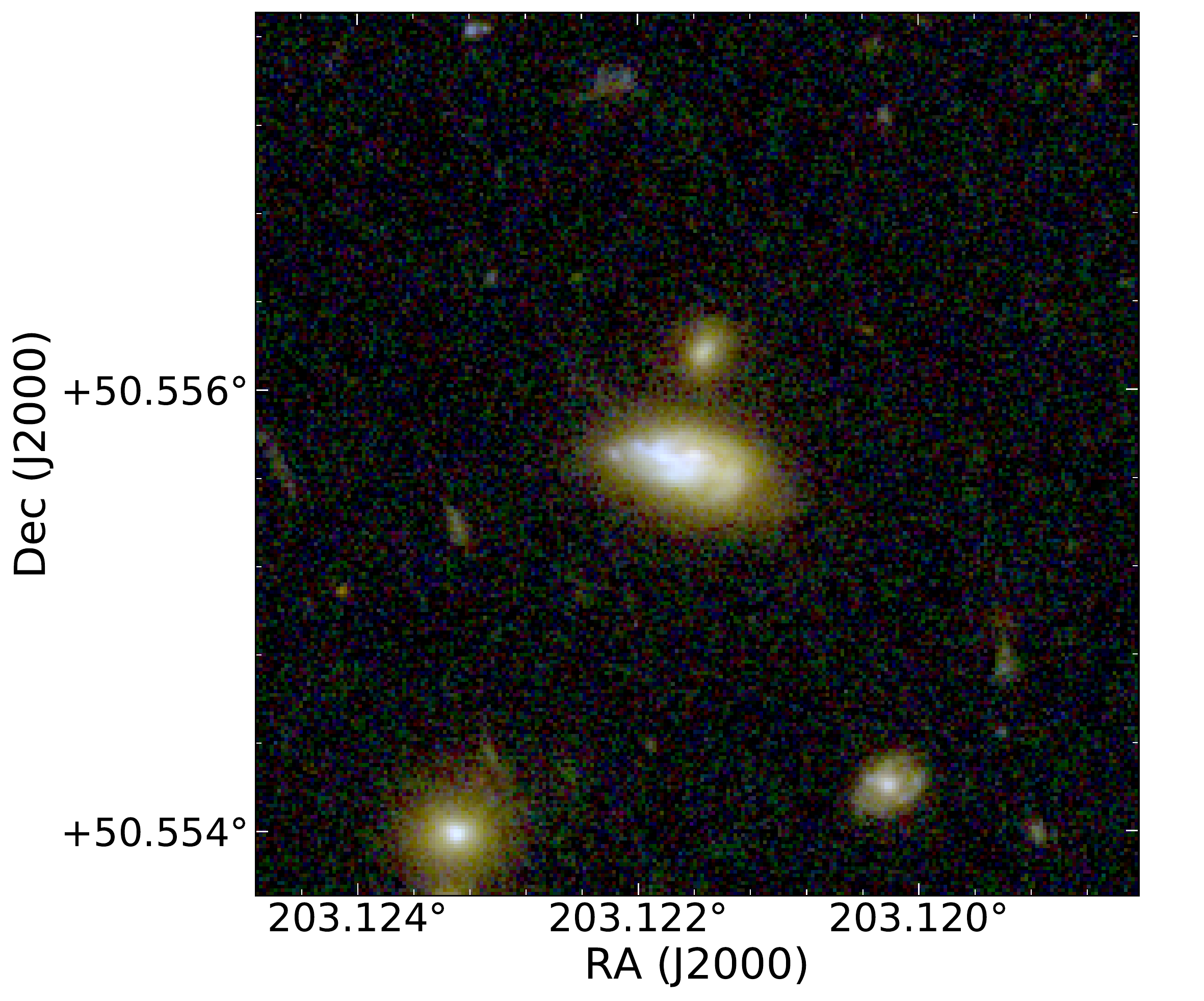}\includegraphics[width=0.32\linewidth, keepaspectratio]{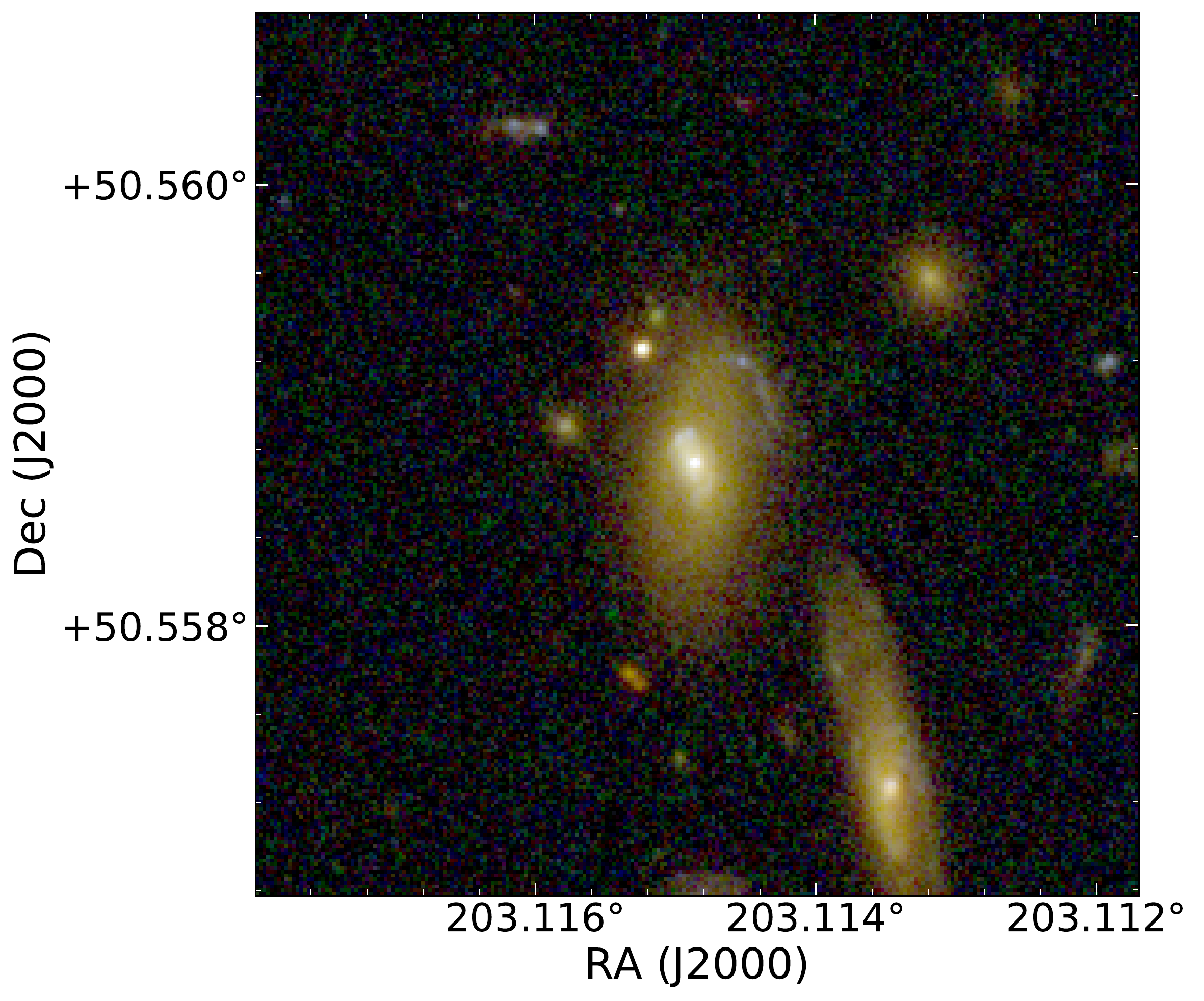}\includegraphics[width=0.32\linewidth, keepaspectratio]{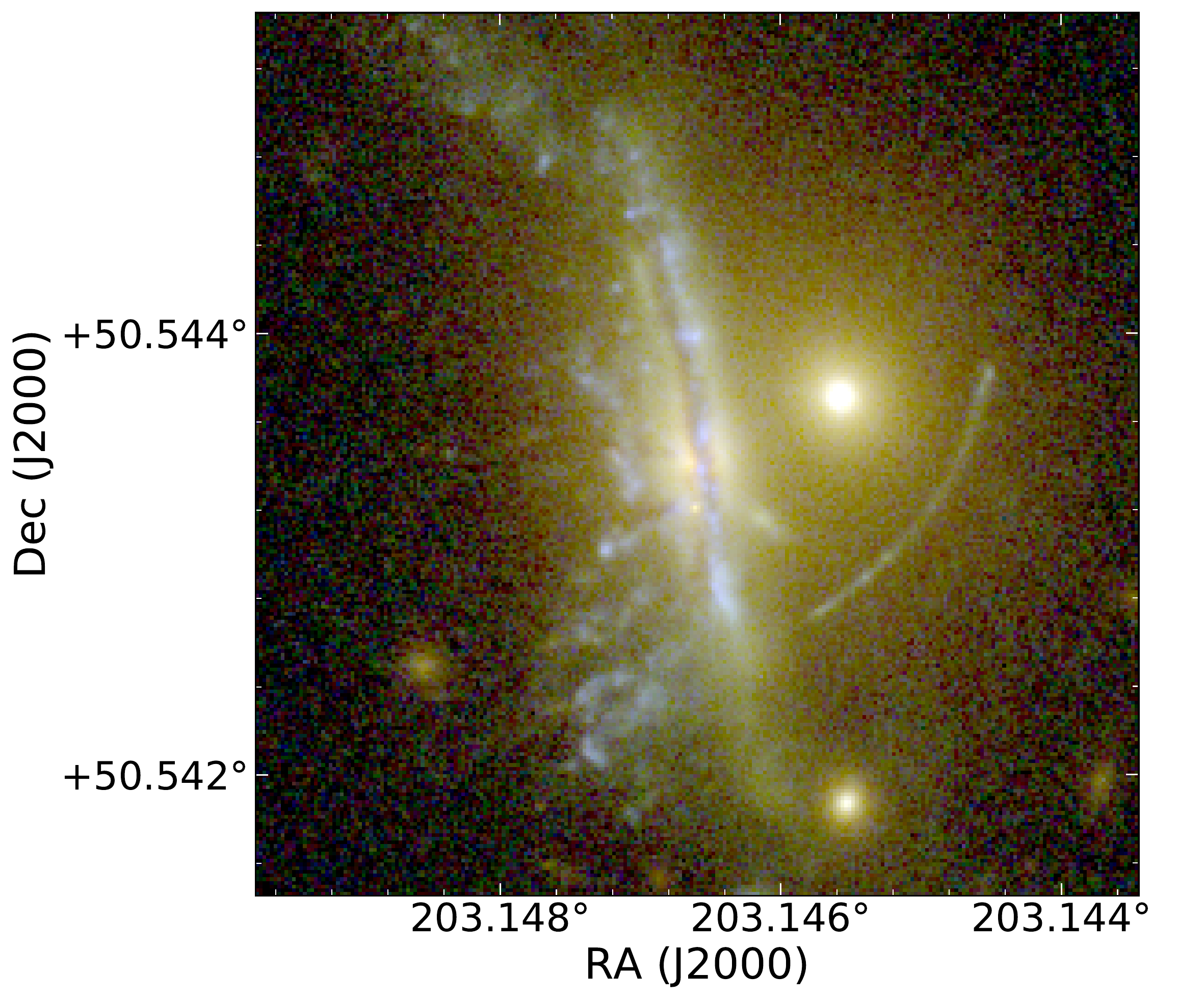}

\includegraphics[width=0.32\linewidth, keepaspectratio]{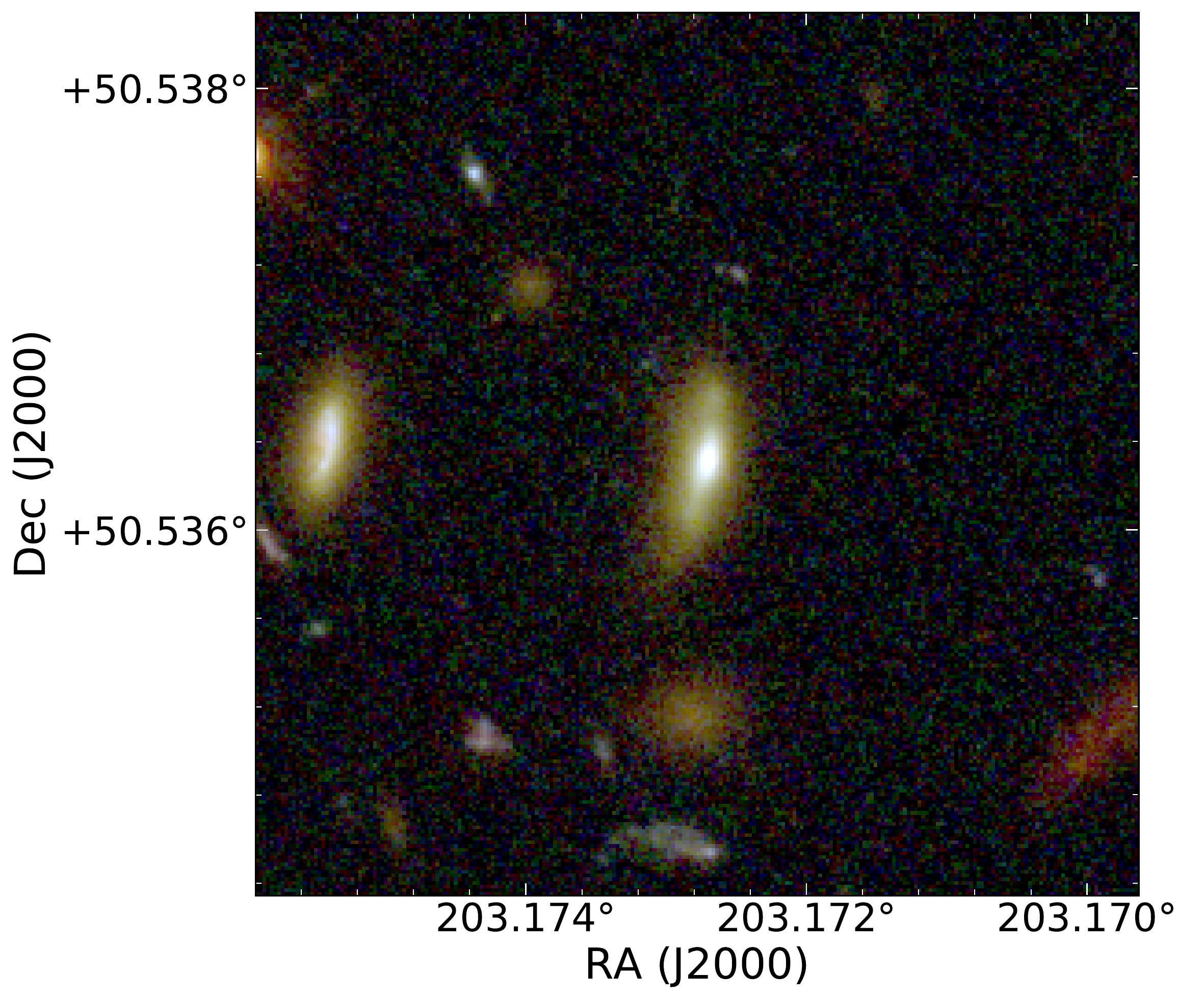}\includegraphics[width=0.32\linewidth, keepaspectratio]{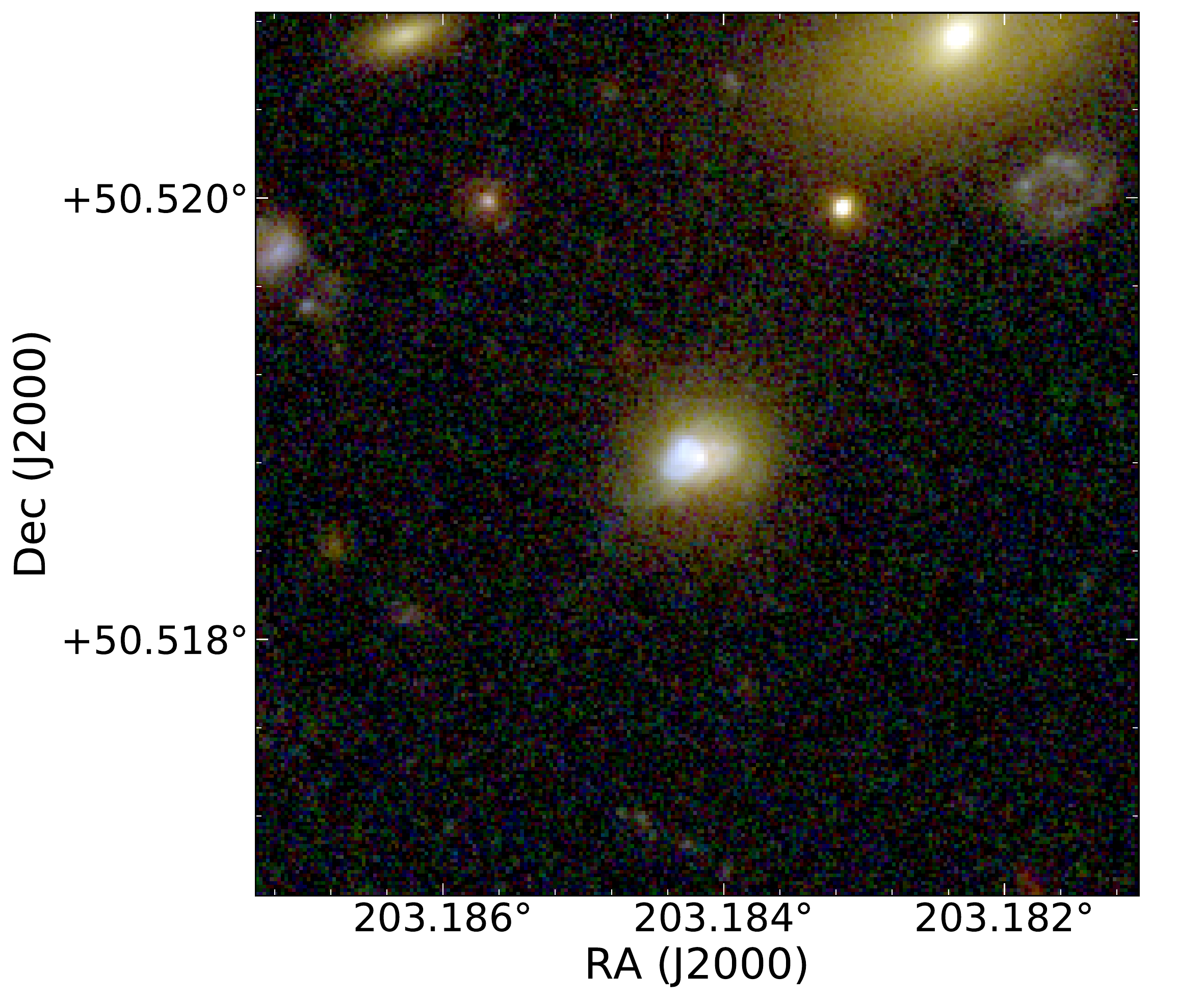}\includegraphics[width=0.32\linewidth, keepaspectratio]{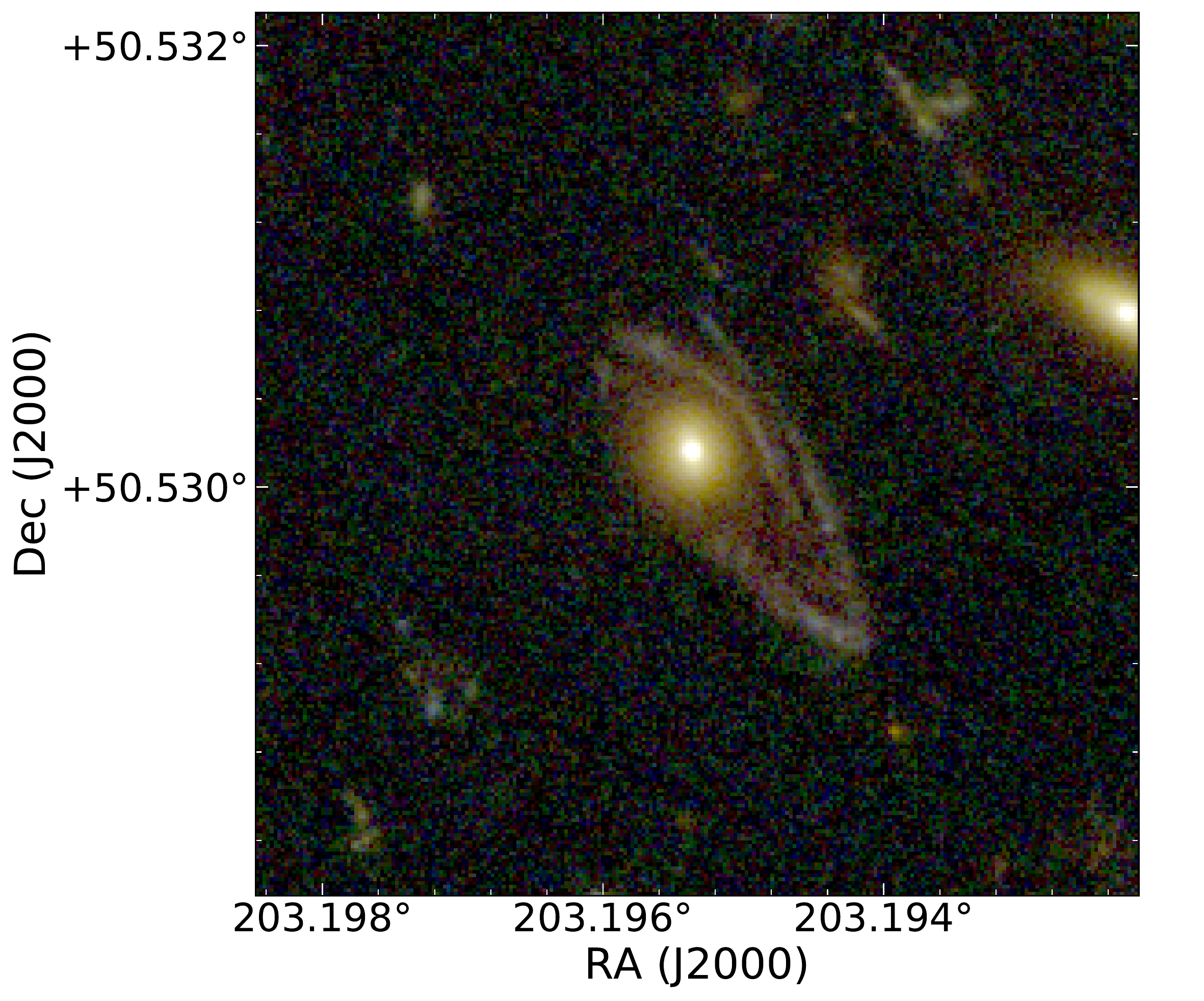}

\includegraphics[width=0.32\linewidth, keepaspectratio]{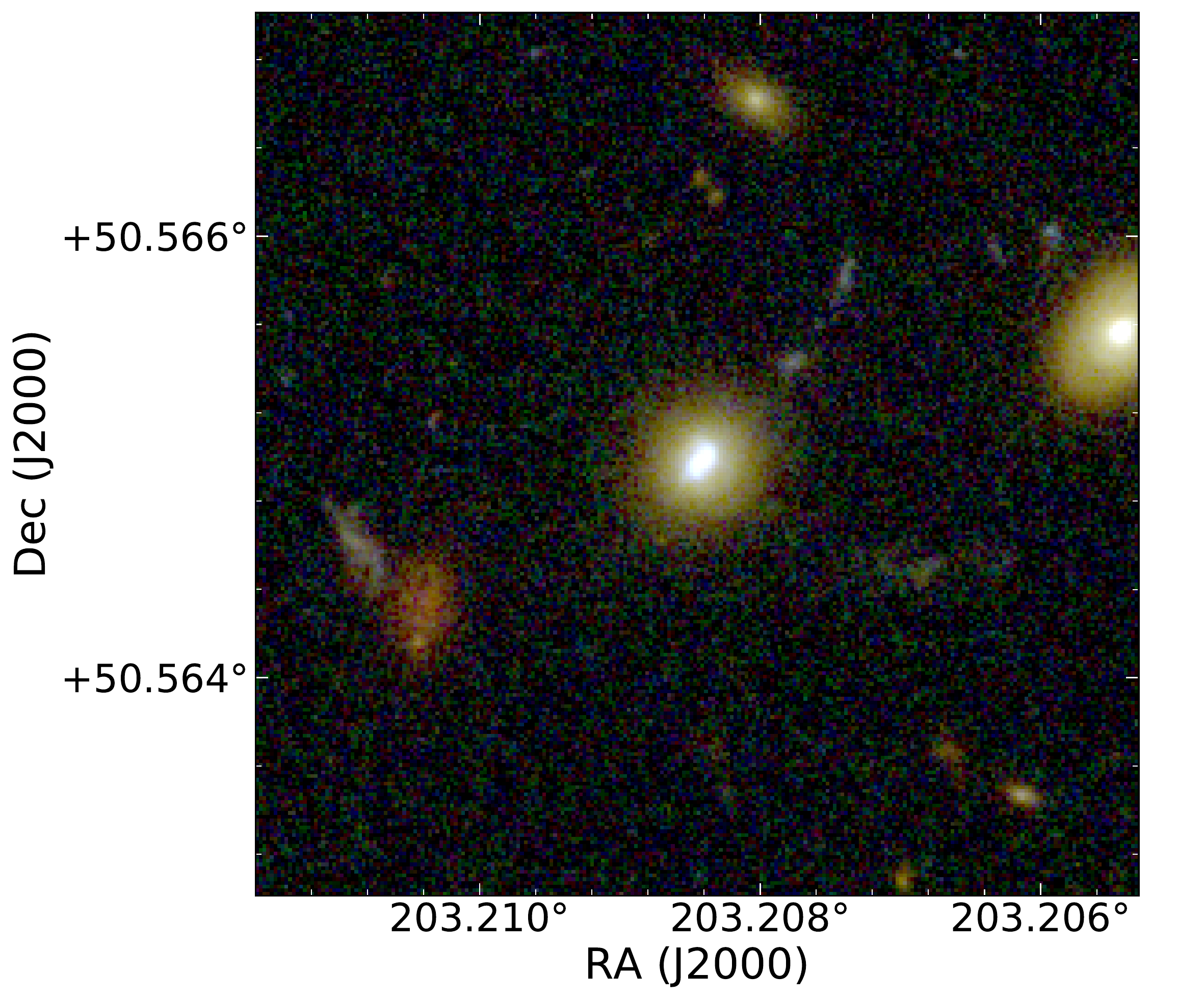}\includegraphics[width=0.32\linewidth, keepaspectratio]{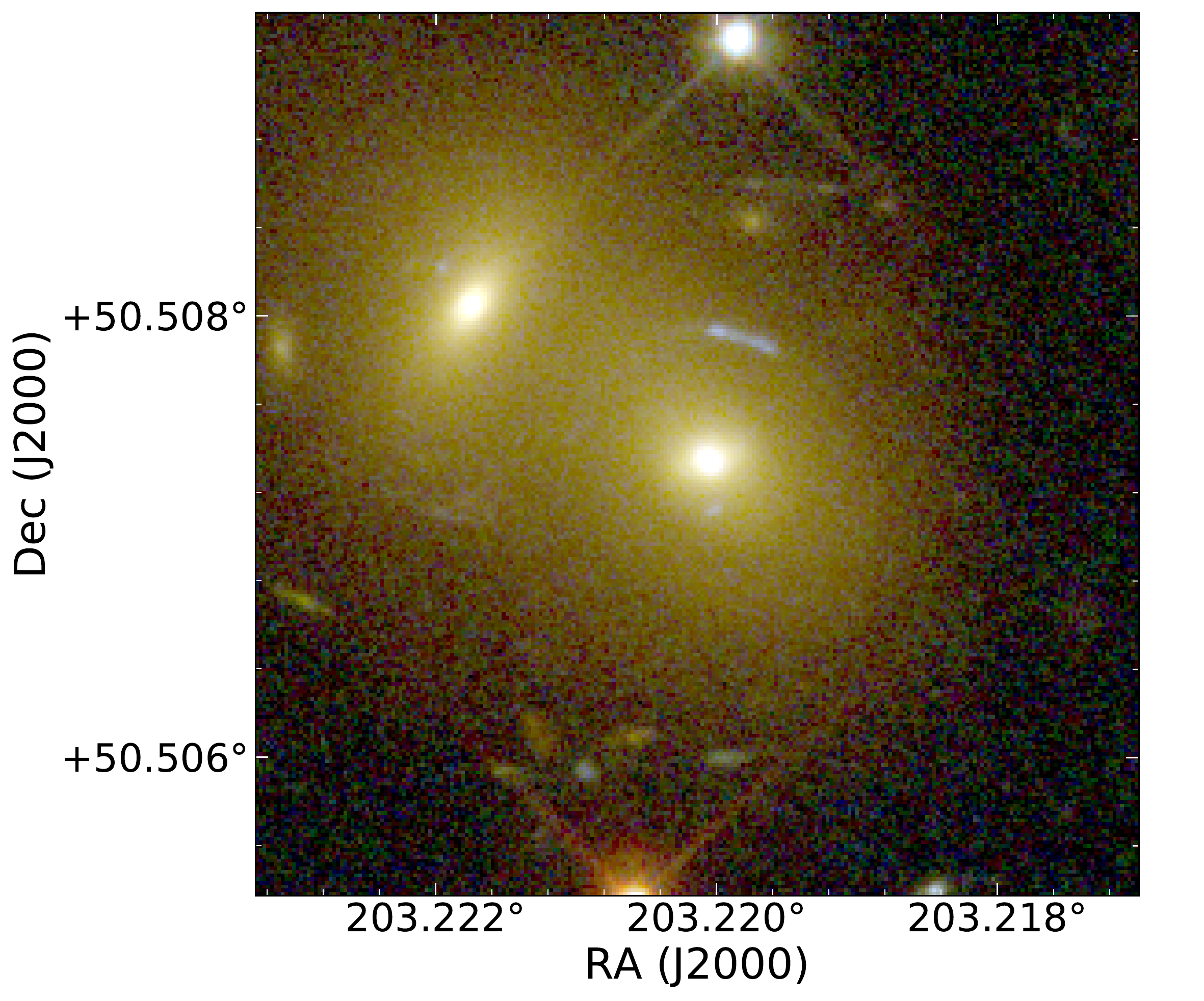}\includegraphics[width=0.32\linewidth, keepaspectratio]{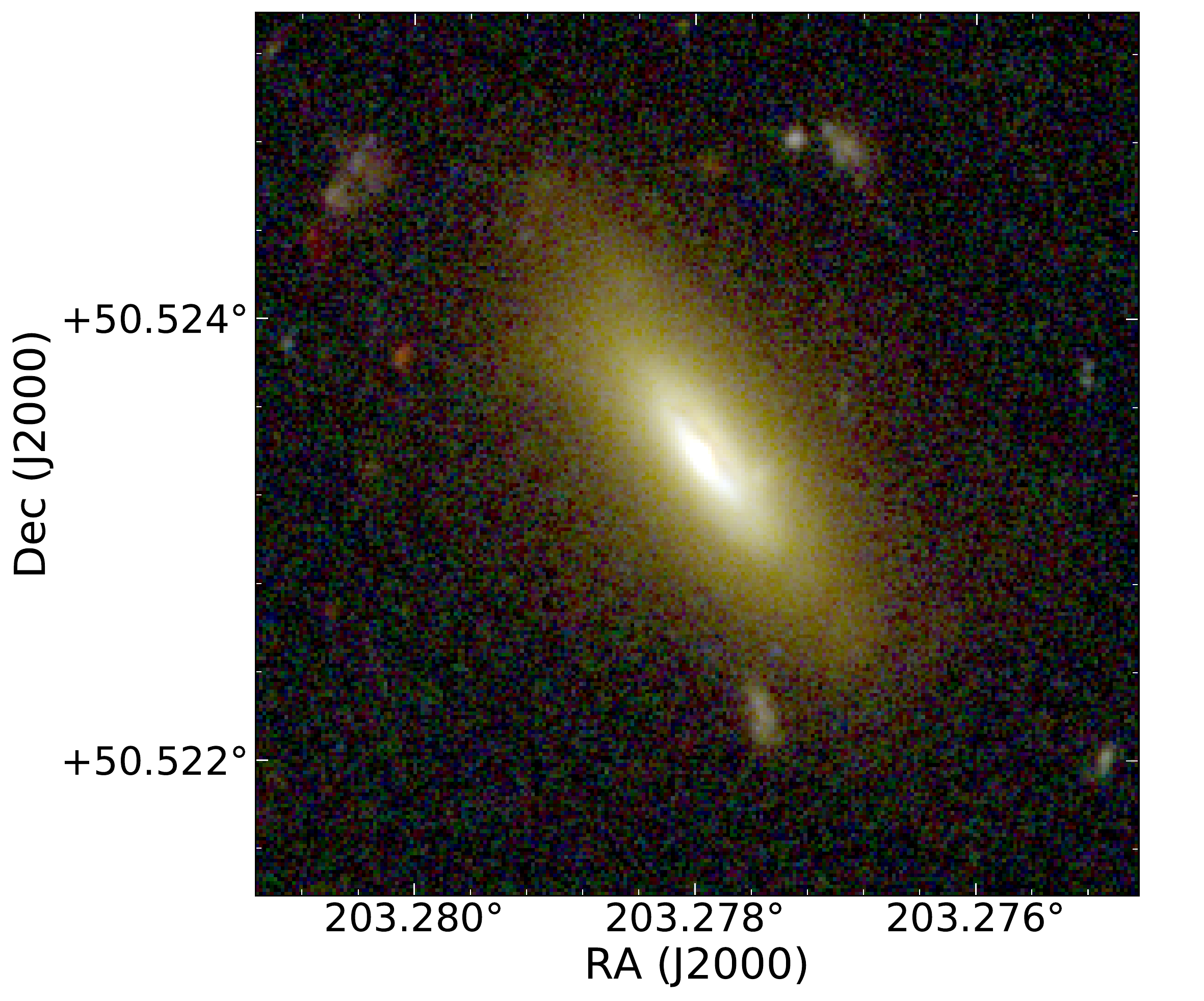}
\caption{Postage stamps ($\approx$ 30 kpc in radius) centered on star-forming cluster members from the HST-ACS RELICS survey \citep{coe19}, in F435W, F606W and F814W filters. The top-right galaxy is the ram-pressure stripping candidate from \citet{ebeling19}.}\label{relics_cuts}
\end{figure*}
MB, GPS, and SLM acknowledge support from the Science and Technology Facilities Council through grant number ST/N021702/1. CPH acknowledges the hospitality of the University of Birmingham while part of this work was completed. MB acknowledges Francesco Calura for the helpful discussions and for providing the theoretical evolutionary tracks from \citet{calura16}.

\appendix

\section{Physical properties of star-forming galaxies}
In this section, we include additional postage stamps of star-forming cluster members covered by HST-ACS imaging from the RELICS survey \citep{coe19} and tables with RA, Dec, redshift, $\rm Log(M_*)$, $\rm Log(M_*)$, $\rm Log(L_{IR})$, $\rm Log(M_{dust})$, $\rm Log(SFR)$ of both cluster and field star-forming galaxies.

\clearpage
\onecolumn

\begin{longtable}{c c c c c c c }
\caption{Star-forming cluster galaxies' properties from SED fitting. From left to right: RA, Dec, redshift, $\rm Log(M_*)$, $\rm Log(L_{IR})$, $\rm Log(M_{dust})$, $\rm Log(SFR)$.}\label{sed_prop_c}\\
\hline \hline
RA(J2000) & Dec(J2000) & Redshift & $\rm Log(M_*\, [M_{\odot}) ]$ &  $\rm Log(L_{IR}\, [L_{\odot}) ]$ &  $\rm Log(M_{dust}\, [M_{\odot}) ]$ &  $\rm Log(SFR\, [M_{\odot}\,yr^{-1}) ]$  \\
\hline \hline 
203.12963 & 50.33531 & 0.27110 & 10.391 & 10.908 & 7.666 & 0.494 \\
203.12941 & 50.45387 & 0.27095 & 10.940 & 10.199 & 6.848 & 0.141 \\
203.12189 & 50.35836 & 0.27283 & 10.027 & 10.563 & 7.979 & 0.631 \\
203.11398 & 50.47497 & 0.27763 & 10.539 & 11.157 & 8.047 & 0.865 \\
203.11030 & 50.37495 & 0.27199 & 10.800 & 10.271 & 8.002 & -0.202 \\
203.09656 & 50.46001 & 0.27091 & 10.323 & 10.151 & 6.871 & 0.075 \\
203.08867 & 50.50301 & 0.28861 & 10.830 & 10.061 & 5.738 & 0.347 \\
203.07966 & 50.37811 & 0.28005 & 9.698 & 10.305 & 7.695 & 0.211 \\
203.06505 & 50.48767 & 0.27746 & 10.484 & 10.598 & 7.054 & 0.481 \\
203.06269 & 50.40175 & 0.28260 & 11.080 & 10.777 & 8.710 & 0.408 \\
203.05391 & 50.40992 & 0.28645 & 9.272 & 10.730 & 7.747 & 0.707 \\
203.04403 & 50.46844 & 0.27967 & 10.983 & 10.740 & 8.478 & 0.519 \\
203.03925 & 50.50684 & 0.27926 & 10.551 & 10.762 & 8.434 & 0.534 \\
203.01853 & 50.36170 & 0.26850 & 9.166 & 10.516 & 7.577 & 0.625 \\
203.00579 & 50.46699 & 0.27801 & 11.300 & 10.422 & 8.310 & -0.813 \\
202.87102 & 50.45685 & 0.27339 & 10.307 & 10.155 & 7.589 & -0.065 \\
202.95994 & 50.34179 & 0.27543 & 10.806 & 10.478 & 7.842 & 0.643 \\
203.49093 & 50.58055 & 0.27081 & 10.601 & 10.180 & 6.724 & -0.012 \\
203.53539 & 50.62541 & 0.27015 & 10.244 & 10.268 & 7.386 & -0.034 \\
203.55738 & 50.59745 & 0.27091 & 9.612 & 10.355 & 7.376 & 0.046 \\
203.09154 & 50.51471 & 0.26978 & 10.457 & 10.283 & 7.018 & -0.522 \\
203.06382 & 50.51408 & 0.28067 & 10.380 & 10.361 & 7.081 & 0.041 \\
203.01598 & 50.52427 & 0.27143 & 10.900 & 10.269 & 6.835 & -0.525 \\
203.07623 & 50.52762 & 0.27927 & 10.085 & 10.512 & 6.355 & -0.094 \\
202.92631 & 50.54256 & 0.27991 & 10.698 & 10.541 & 7.548 & 0.270 \\
202.96385 & 50.54498 & 0.27954 & 10.255 & 10.795 & 7.619 & 0.104 \\
203.12158 & 50.55571 & 0.28307 & 9.751 & 10.536 & 7.796 & 0.412 \\
203.11484 & 50.55878 & 0.28292 & 10.425 & 10.238 & 7.992 & -0.024 \\
203.11450 & 50.58810 & 0.27387 & 10.203 & 10.785 & 7.669 & 0.436 \\
203.06396 & 50.58839 & 0.28253 & 9.990 & 10.261 & 6.368 & -0.218 \\
203.00356 & 50.59170 & 0.28496 & 10.495 & 10.652 & 7.349 & 0.395 \\
203.01275 & 50.59240 & 0.27623 & 10.204 & 10.984 & 8.082 & 0.824 \\
202.90261 & 50.59530 & 0.27387 & 10.574 & 10.855 & 8.265 & 0.453 \\
202.93216 & 50.70218 & 0.27619 & 10.537 & 10.753 & 8.590 & 0.419 \\
202.92527 & 50.69716 & 0.28016 & 10.594 & 10.860 & 8.674 & 0.569 \\
203.08024 & 50.60107 & 0.28469 & 10.441 & 10.542 & 7.575 & 0.186 \\
203.12514 & 50.68257 & 0.27703 & 10.487 & 10.793 & 7.817 & 1.345 \\
203.12229 & 50.66669 & 0.28257 & 10.433 & 10.973 & 8.323 & 0.282 \\
203.09918 & 50.66675 & 0.27768 & 10.266 & 10.103 & 7.920 & -0.322 \\
203.13003 & 50.66507 & 0.28091 & 10.721 & 11.202 & 8.679 & 0.970 \\
203.08016 & 50.66446 & 0.28209 & 10.036 & 10.081 & 7.003 & -0.030 \\
202.86317 & 50.65987 & 0.27462 & 9.818 & 10.415 & 7.346 & 0.263 \\
203.10397 & 50.65570 & 0.27736 & 9.820 & 10.605 & 8.076 & 0.481 \\
203.10819 & 50.63843 & 0.28307 & 10.077 & 10.389 & 8.304 & -0.140 \\
202.99090 & 50.62927 & 0.28579 & 11.225 & 10.347 & 7.292 & -0.888 \\
203.08126 & 50.63134 & 0.28396 & 9.982 & 10.241 & 8.031 & 0.231 \\
202.87612 & 50.61704 & 0.27357 & 9.631 & 10.214 & 6.782 & 0.222 \\
203.25648 & 50.27521 & 0.27551 & 10.002 & 10.040 & 6.161 & 0.146 \\
203.29989 & 50.25200 & 0.27560 & 10.782 & 10.533 & 8.463 & 0.185 \\
203.14659 & 50.54345 & 0.27245 & 10.728 & 10.643 & 8.307 & 0.867 \\
203.14596 & 50.59712 & 0.29223 & 10.249 & 10.339 & 6.983 & 0.284 \\
203.14467 & 50.52873 & 0.28796 & 10.041 & 10.728 & 7.994 & 0.514 \\
203.17298 & 50.69662 & 0.27727 & 9.928 & 11.029 & 8.396 & 0.916 \\
203.17278 & 50.53634 & 0.26867 & 9.351 & 10.058 & 6.695 & 0.071 \\
203.18419 & 50.51884 & 0.28733 & 9.501 & 10.646 & 8.332 & 0.551 \\
203.19533 & 50.53015 & 0.27962 & 10.416 & 10.667 & 8.398 & 0.233 \\
203.21420 & 50.55136 & 0.28499 & 11.039 & 11.031 & 8.731 & 0.920 \\
203.20845 & 50.56501 & 0.27216 & 9.438 & 10.467 & 7.880 & 0.527 \\
203.20933 & 50.62268 & 0.27641 & 9.400 & 10.227 & 6.251 & 0.336 \\
203.21614 & 50.61397 & 0.28869 & 11.326 & 10.576 & 8.433 & -0.397 \\
203.22014 & 50.50737 & 0.28072 & 11.220 & 10.238 & 7.501 & -0.587 \\
203.22470 & 50.53909 & 0.26592 & 10.715 & 10.750 & 7.873 & 0.465 \\
203.24412 & 50.65999 & 0.27289 & 10.935 & 10.174 & 7.563 & 0.099 \\
203.27799 & 50.52339 & 0.28521 & 10.902 & 10.219 & 6.862 & -0.466 \\
203.47682 & 50.59392 & 0.26896 & 10.655 & 10.679 & 7.522 & 0.567 \\
203.47320 & 50.57590 & 0.27184 & 11.057 & 10.344 & 6.991 & 0.405 \\
203.47313 & 50.61701 & 0.26945 & 10.329 & 10.070 & 6.434 & -0.439 \\
203.37712 & 50.56861 & 0.26912 & 10.153 & 10.676 & 7.989 & 0.808 \\
203.44903 & 50.65999 & 0.27299 & 10.680 & 9.840 & 6.383 & -0.332 \\
203.36335 & 50.50971 & 0.27526 & 10.168 & 10.474 & 8.112 & 1.025 \\
203.28109 & 50.57133 & 0.28197 & 9.625 & 10.524 & 8.261 & 0.398 \\
203.34024 & 50.56264 & 0.28453 & 10.240 & 10.165 & 6.504 & 0.046 \\
203.29700 & 50.52530 & 0.27935 & 10.209 & 10.331 & 7.720 & 0.244 \\
203.29734 & 50.56141 & 0.27204 & 10.134 & 10.891 & 7.844 & 0.629 \\
203.30944 & 50.54776 & 0.26924 & 10.067 & 10.160 & 7.505 & -0.194 \\
202.74531 & 50.51142 & 0.27869 & 10.903 & 10.659 & 7.490 & 0.259 \\
202.77619 & 50.54283 & 0.27376 & 10.387 & 10.013 & 7.336 & -0.144 \\
202.72388 & 50.40450 & 0.27296 & 10.249 & 10.482 & 7.171 & 0.407 \\
203.29128 & 50.48398 & 0.27714 & 10.350 & 10.297 & 8.018 & -0.417 \\
203.20539 & 50.48067 & 0.27680 & 10.077 & 10.274 & 7.478 & 0.090 \\
203.44439 & 50.47987 & 0.27542 & 9.922 & 10.929 & 8.071 & 0.896 \\
203.29094 & 50.47598 & 0.29022 & 9.879 & 10.353 & 6.626 & 0.411 \\
203.21088 & 50.46454 & 0.26757 & 10.281 & 10.437 & 7.084 & 0.463 \\
203.22324 & 50.46001 & 0.27721 & 10.310 & 10.537 & 7.300 & -0.009 \\
203.23358 & 50.43712 & 0.28032 & 10.651 & 10.301 & 7.379 & -0.057 \\
203.21713 & 50.41843 & 0.27916 & 10.635 & 10.806 & 7.839 & 0.515 \\
203.19079 & 50.41370 & 0.26762 & 10.265 & 10.715 & 7.960 & 0.414 \\
203.15416 & 50.32823 & 0.27409 & 10.559 & 10.735 & 8.279 & 0.595 \\
203.38212 & 50.38065 & 0.26707 & 10.032 & 11.429 & 8.213 & 1.323 \\
203.19602 & 50.39308 & 0.28044 & 10.888 & 10.201 & 7.914 & -0.568 \\
\end{longtable}

\begin{longtable}{c c c c c c c }
\caption{Star-forming field galaxies' properties from SED fitting. From left to right:  RA, Dec, redshift, $\rm Log(M_*)$, $\rm Log(L_{IR})$, $\rm Log(M_{dust})$, $\rm Log(SFR)$.}\label{sed_prop_f}\\
\hline \hline
RA(J2000) & Dec(J2000) & Redshift & $\rm Log(M_*\, [M_{\odot}) ]$ &  $\rm Log(L_{IR}\, [L_{\odot}) ]$ &  $\rm Log(M_{dust}\, [M_{\odot}) ]$ &  $\rm Log(SFR\, [M_{\odot}\,yr^{-1}) ]$  \\
\hline \hline 
203.12549 & 50.49451 & 0.23277 & 9.841 & 10.021 & 6.997 & -0.072 \\
202.91964 & 50.43139 & 0.24818 & 10.672 & 10.453 & 7.494 & -0.253 \\
202.96486 & 50.38590 & 0.23292 & 11.064 & 10.661 & 7.959 & 0.157 \\
203.12311 & 50.59055 & 0.24985 & 10.234 & 10.553 & 8.213 & 0.256 \\
203.08076 & 50.71770 & 0.24995 & 10.571 & 10.774 & 7.678 & 0.508 \\
203.01838 & 50.70767 & 0.24926 & 9.682 & 10.485 & 7.234 & 0.566 \\
203.10381 & 50.62614 & 0.29538 & 9.816 & 9.881 & 6.846 & -0.445 \\
203.14128 & 50.71727 & 0.24950 & 11.217 & 11.474 & 8.542 & 1.112 \\
203.44016 & 50.61321 & 0.26022 & 10.311 & 10.293 & 8.111 & -0.064 \\
203.39635 & 50.47473 & 0.24962 & 10.249 & 10.918 & 8.184 & 0.717 \\
203.42308 & 50.42994 & 0.24905 & 10.454 & 10.750 & 7.808 & 0.532 \\
203.35097 & 50.41664 & 0.24969 & 10.091 & 11.015 & 8.294 & 0.985 \\
203.20135 & 50.36860 & 0.25973 & 10.335 & 10.545 & 7.737 & -0.070 \\
203.41938 & 50.38865 & 0.23322 & 10.449 & 11.118 & 8.579 & 0.917 \\
250.05125 & 46.77683 & 0.27764 & 10.325 & 10.718 & 7.955 & 0.266 \\
250.07478 & 46.88606 & 0.27239 & 10.174 & 10.362 & 7.329 & 0.037 \\
249.97753 & 46.88248 & 0.26509 & 10.116 & 10.871 & 7.714 & 0.636 \\
249.93895 & 46.85836 & 0.25392 & 10.034 & 10.656 & 8.320 & 0.501 \\
249.85594 & 46.83027 & 0.28305 & 10.129 & 11.019 & 7.645 & 0.945 \\
250.43362 & 46.79677 & 0.27245 & 9.858 & 11.078 & 7.821 & 0.582 \\
250.44185 & 46.76573 & 0.25299 & 10.565 & 11.162 & 7.950 & 1.078 \\
250.06733 & 46.58628 & 0.24598 & 10.488 & 10.742 & 8.211 & 0.705 \\
250.01354 & 46.59893 & 0.26302 & 10.831 & 10.809 & 7.788 & 0.547 \\
250.29164 & 46.65747 & 0.26260 & 10.759 & 11.213 & 8.393 & 0.885 \\
250.15680 & 46.64951 & 0.26406 & 10.313 & 11.061 & 8.007 & 0.574 \\
250.38557 & 46.63302 & 0.26051 & 9.765 & 9.803 & 5.910 & -0.038 \\
250.22619 & 46.63078 & 0.26244 & 10.149 & 10.702 & 8.543 & 0.445 \\
250.12356 & 46.49810 & 0.26205 & 10.474 & 11.497 & 8.029 & 1.321 \\
250.15085 & 46.52782 & 0.25423 & 10.561 & 10.833 & 7.826 & 0.612 \\
250.04227 & 46.48227 & 0.26596 & 11.173 & 10.485 & 7.971 & 0.231 \\
250.17092 & 46.45124 & 0.28477 & 9.633 & 11.004 & 8.421 & 0.990 \\
250.17490 & 46.42400 & 0.26076 & 10.289 & 10.223 & 7.476 & 0.274 \\
322.45662 & 0.09840 & 0.28836 & 10.279 & 11.016 & 7.651 & 0.904 \\
322.22652 & -0.01701 & 0.29040 & 10.621 & 11.161 & 8.450 & 0.470 \\
322.23550 & 0.00328 & 0.28951 & 10.749 & 10.490 & 7.632 & 0.576 \\
322.26192 & -0.07151 & 0.28237 & 9.476 & 10.604 & 7.961 & 0.365 \\
322.31268 & -0.07008 & 0.28984 & 10.385 & 10.400 & 7.178 & 0.375 \\
322.55007 & 0.05568 & 0.27415 & 9.651 & 10.628 & 8.017 & 0.553 \\
322.45448 & -0.00827 & 0.28122 & 10.512 & 11.333 & 7.641 & 0.477 \\
126.70602 & 4.25056 & 0.28282 & 10.798 & 10.579 & 7.833 & 0.488 \\
126.68767 & 4.26693 & 0.27627 & 10.080 & 11.476 & 7.971 & 1.398 \\
126.58493 & 4.44359 & 0.28403 & 10.559 & 10.530 & 6.961 & 0.492 \\
126.68624 & 4.15960 & 0.27666 & 10.761 & 10.148 & 6.869 & 0.359 \\
126.68559 & 4.11816 & 0.26170 & 11.057 & 11.160 & 7.682 & 0.709 \\
135.14587 & 21.10720 & 0.28624 & 10.347 & 10.227 & 5.582 & 0.338 \\
135.04694 & 21.02492 & 0.27581 & 10.924 & 11.098 & 7.957 & 0.905 \\
135.37527 & 20.93561 & 0.27862 & 9.308 & 10.450 & 6.813 & 0.506 \\
203.95883 & 41.22924 & 0.26982 & 10.875 & 10.989 & 8.777 & 0.725 \\
203.53119 & 41.15471 & 0.29399 & 10.239 & 10.249 & 7.895 & -0.171 \\
203.77227 & 41.04787 & 0.25878 & 10.390 & 10.630 & 7.326 & 0.279 \\
203.55147 & 41.08073 & 0.25945 & 9.709 & 10.616 & 7.946 & -0.384 \\
203.65717 & 41.17331 & 0.29746 & 10.562 & 11.307 & 7.853 & 1.160 \\
203.66612 & 41.16584 & 0.29690 & 10.942 & 10.566 & 7.756 & -0.171 \\
203.79205 & 41.14138 & 0.29659 & 10.390 & 10.581 & 7.740 & 0.406 \\
203.86673 & 41.00258 & 0.29197 & 10.768 & 10.109 & 6.448 & 0.111 \\
203.93678 & 41.16624 & 0.26600 & 9.926 & 11.084 & 8.435 & 0.951 \\
204.07297 & 41.08957 & 0.29784 & 10.758 & 10.563 & 8.183 & 0.240 \\
204.05708 & 41.04166 & 0.29757 & 10.232 & 10.660 & 7.913 & 0.054 \\
203.82568 & 40.84622 & 0.27306 & 9.926 & 10.232 & 6.825 & 0.783 \\
203.58752 & 40.95970 & 0.27346 & 10.196 & 10.684 & 8.348 & 0.510 \\
204.02733 & 40.97041 & 0.29209 & 10.141 & 11.756 & 8.250 & 0.955 \\
204.03830 & 40.94977 & 0.29686 & 9.821 & 10.528 & 8.392 & 0.540 \\
203.96866 & 40.95030 & 0.29159 & 10.080 & 10.115 & 7.368 & 0.154 \\
204.01956 & 40.93465 & 0.29769 & 9.766 & 11.048 & 7.793 & 0.983 \\
203.97250 & 40.91645 & 0.28238 & 10.827 & 11.255 & 8.665 & 0.649 \\
204.11808 & 40.84244 & 0.29926 & 10.206 & 10.611 & 8.324 & 0.660 \\
204.10542 & 40.85266 & 0.29958 & 10.283 & 10.737 & 7.900 & 0.751 \\
204.04023 & 40.86997 & 0.27133 & 9.622 & 10.757 & 8.544 & 0.717 \\
\end{longtable}
\twocolumn

\bibliographystyle{mnras}
\bibliography{biblio.bib}

\bsp	
\label{lastpage}

\end{document}